\documentclass[a4paper,pre,reqno,superscriptaddress,twocolumn,showpacs]{revtex4-1}

\usepackage{graphicx}
\usepackage{dcolumn}
\usepackage{epsfig}

\usepackage[centertags]{amsmath}
\usepackage{amsfonts,amsmath}
\usepackage[mathscr]{euscript}
\usepackage{amssymb}
\usepackage{amsthm}
\usepackage{newlfont}
\usepackage{mathrsfs}
\usepackage{subfigure}

\usepackage{changes}

\newcommand{\opunit}{\text{1}\kern-0.22em\text{l}}

\usepackage{color}
\definecolor{dgreen}{rgb}{0,0.7,0}

\newcommand{\ie}{\textit{i.e.}}

\newcommand{\id}{\textrm{d}}

\def\bea{\begin{eqnarray}}
\def\eea{\end{eqnarray}}
\def\ba{\begin{array}}
\def\ea{\end{array}}
\def\n{\nonumber}
\def\la{\langle}
\def\ra{\rangle}

\def\Jd{J_\text{d}}
\def \Jres {J_\text{reset}}

\begin{document}

\title{Symmetric Exclusion Process  under Stochastic Resetting}
%\author{Urna Basu, Anupam Kundu, Arnab Pal}
\author{Urna Basu}
\affiliation{Raman Research Institute, Bengaluru 560080, India}
\author{Anupam Kundu}
\affiliation{International Centre for Theoretical Sciences, Tata Institute of Fundamental Research, Bengaluru 560089, India}
\author{Arnab Pal}
\affiliation{School of Chemistry, Raymond and Beverly Sackler Faculty of Exact Sciences, Tel Aviv University, Tel Aviv 6997801, Israel}
\affiliation{Center for the Physics and Chemistry of Living Systems. Tel Aviv University, 6997801, Tel Aviv, Israel}
\affiliation{The Sackler Center for Computational Molecular and Materials Science, Tel Aviv University, 6997801, Tel Aviv, Israel}

\begin{abstract}

We study the behaviour of a Symmetric Exclusion Process (SEP) in presence of stochastic resetting where the configuration of the system is reset to a step-like profile with a fixed rate $r.$ We show that the presence of resetting affects both the stationary and dynamical properties of SEP strongly. We compute the exact time-dependent density profile and show that the stationary state is characterized by a non-trivial inhomogeneous profile in contrast to the flat one for $r=0.$  We also show that for $r>0$ the average diffusive current grows linearly with time $t,$ in stark contrast to the $\sqrt{t}$ growth for $r=0.$ In addition to the underlying diffusive current, we identify the resetting current in the system which emerges due to the sudden relocation of the particles to the step-like configuration and is strongly correlated to the diffusive current. We show that the average resetting current is negative, but its magnitude also grows linearly with time $t.$ We also compute the probability distributions of the diffusive current, resetting current and the total current (sum of the diffusive and the resetting currents) using the renewal approach. We demonstrate that while the typical fluctuations of both the diffusive and reset currents around the mean are typically Gaussian, the distribution of the total current shows a strong non-Gaussian behaviour. 

\end{abstract}

\maketitle

\section{Introduction}

Stochastic resetting, which refers to intermittent interruption and restart of a dynamical process, has been a subject of immense interest in recent years. It has found applications in a wide range of areas starting from search problems \cite{search1,search4,Arnab2017,Arnab2019}, population dynamics \cite{population1, population2}, enzymatic catalysis \cite{catalysis,bio4} to computer science\cite{search2,search3}, stock markets \cite{stock} and biological processes \cite{bio1, bio2, bio3}. Stochastic resetting of a single Brownian particle is the paradigmatic example where the position of the particle is reset to a fixed point in space with a certain rate \cite{Brownian}. This simple act has drastic consequences on the statistical properties of the particle --- it results in a nontrivial stationary state, anomalous relaxation behaviour, as well as finite mean first passage time.

Several variations and extensions of this simple model have been explored in recent years\cite{Brownian2,Brownian3,absorption,highd,Mendez2016,Puigdellosas,deepak,Arnab2019}.  Specific examples include: resetting in presence of an external potential \cite{Arnab2015, potential}, in a confinement \cite{circle, interval} or to an extended region \cite{CCRW}, and resetting to already excursed positions  \cite{Boyer2014, Sanjib2015}.
Stochastic resetting has also been studied in more general nonequilibrium contexts -- in reaction processes \cite{bio4,catalysis}, L\'{e}vy flights \cite{Levy}, coagulation-diffusion process \cite{coagulation}, telegraphic process \cite{telegraphic}, for run-and-tumble particles \cite{RTP} and to model nonequilibrium baths \cite{Maes2017}. Studies were not only limited to a constant rate resetting, other protocols have also been investigated in great details. These include deterministic resetting \cite{deterministic}, space \cite{Roldan2017} or time dependent \cite{time-dep1,time-dep2} resetting rate, resetting followed by a refractory period\cite{refractory,refractory2}, non-Markovian resetting \cite{ShamikPRE, nonmarkov2, nonmarkov3}  and resetting sensitive to internal dynamics \cite{Evans2017}.

% 
% resetting of underdamped Brownian particle \cite{deepak}, resetting to previously visited positions \cite{Boyer2014}, and resetting to current maximum position \cite{Sanjib2015}. 

% 
%  Additionally, different mechanisms of resetting have also been studied, \eg, resetting with space-dependent rate \cite{Roldan2017}, resetting in an extended region \cite{CCRW}, deterministic resetting \cite{deterministic}, time-dependent resetting \cite{time-dep1,time-dep2}, resetting with refractory period \cite{refractory,refractory2}, non-Markovian resetting protocol \cite{ShamikPRE, non-markov2} and resetting depending on internal dynamics \cite{Evans2017}. However, most of these studies refer to systems with a single degree of freedom. 

\begin{figure}[t]
 \centering
 \includegraphics[width=7.5 cm]{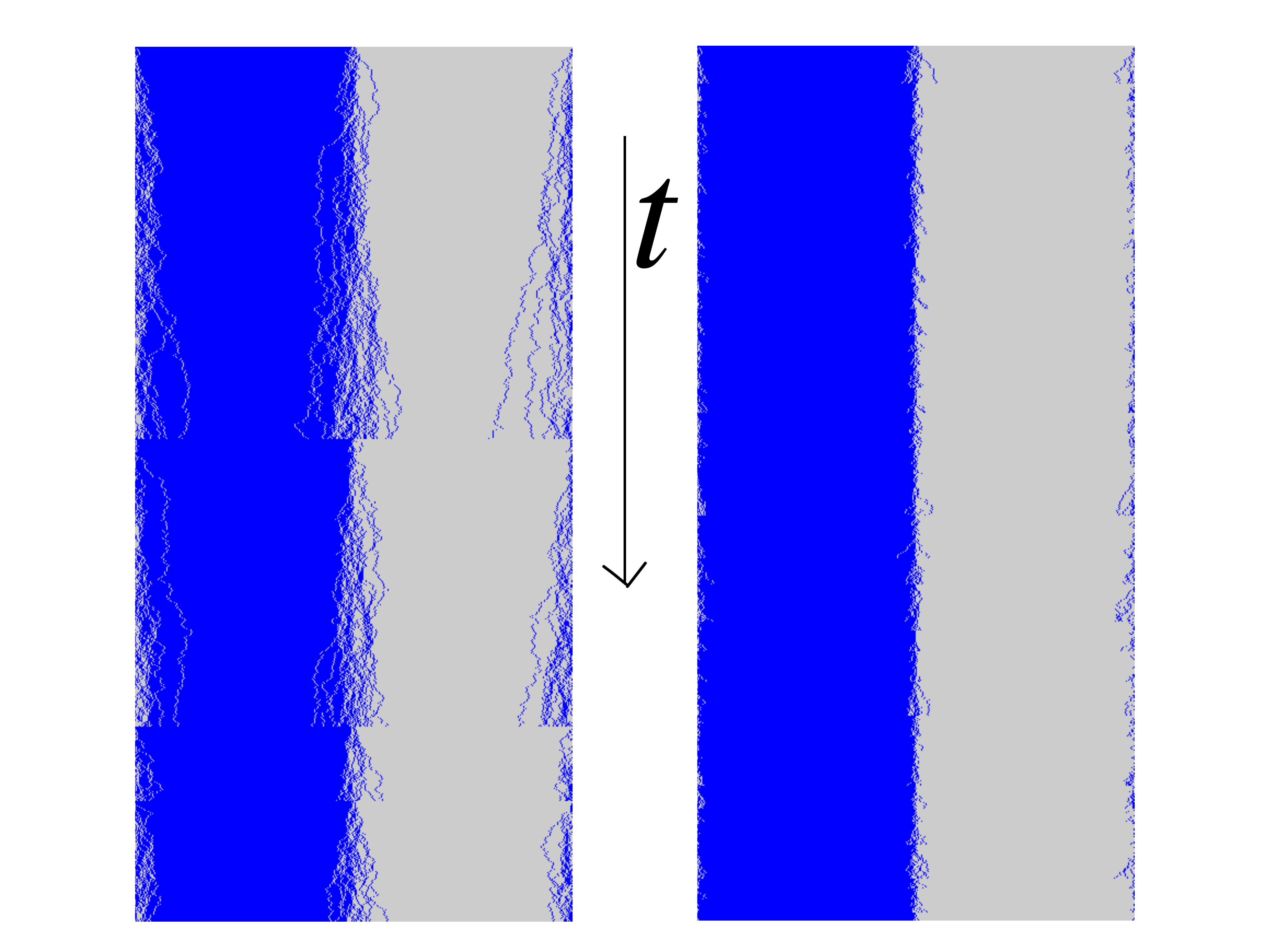}
 % snap.pdf: 0x0 px, 300dpi, 0.00x0.00 cm, bb=
 \caption{Typical snapshots of time evolution of a system of size $L=400$ for two different values of the resetting rate $r=0.01$ (left) and $r=0.1$(right). The dark blue points indicate presence of particles while the light Grey ones indicate empty sites.  For a small value of $r$ (left) the typical duration between two consecutive resetting events are longer, and the particles spread deeper into the empty half of the lattice whereas for larger $r$ (right) the resetting occurs more frequently and the density profile remains close to the step-like initial state. }
 \label{fig:snapshot}
\end{figure}

An important question that naturally arises is how the presence of resetting dynamics affects the systems with many interacting degrees of freedom. This issue has not been explored much so far except for a few handful of  models. These studies include dynamics of KPZ-like fluctuating interfaces\cite{kpz-reset, Shamik2016}, one-dimensional quantum spin chain\cite{KS2018}, and a pair of interacting Brownian particles \cite{Evans2017,Silva2018}. In all these cases, resetting leads to nonequilibrium stationary states, characterized by non-Gaussian fluctuations.  
However, the effect of resetting on the behaviour of current, which plays an important role in characterizing the nonequilibrium stationary state, has not been studied yet. This question is of paramount importance, because presence of stochastic resetting introduces an additional time-scale and is expected to modify the behaviour of current significantly. The  exclusion processes \cite{Liggett}, which are simple well known models of interacting particles, provide a natural playground for exploring these questions.

In this article we study the effect of stochastic resetting on Symmetric Exclusion Process (SEP) \cite{Spitzer, Liggett} and explore how the presence of resetting changes the dynamical and stationary properties of SEP. The stochastic resetting is implemented by interrupting the time-evolution at some rate $r,$ and restarting the process from a specific configuration. It turns out that the incorporation of the resetting mechanism introduces an extra current $\Jres$ in addition to the usual diffusive particle current $J_\text{d}.$
We show that, for $r>0,$ the average diffusive current increases linearly with time $t,$ in contrast to the $\sqrt{t}$ behaviour in the absence of the resetting \cite{Derrida1}. Additionally, the average resetting current also shows a linear temporal growth in magnitude, although it remains negative. We also compute the distribution of the diffusive current $J_\text{d}$, resetting current $\Jres,$ as well as the total current $J_r=J_\text{d}+\Jres$.  We observe that, while the diffusive and resetting current show Gaussian behaviour, the fluctuations of the total current are characterized by a strongly non-Gaussian distribution.

The article is organized as follows: In the next section we define our system and summarize our main results. Sec \ref{sec:density} is devoted to the computation of the time evolution of the density profile under resetting. In Sec. \ref{sec:cur} we investigate the  behaviour of the particle current -- Sec. \ref{sec:J_dif} and \ref{sec:J_reset} focus on the diffusive and resetting currents respectively, whereas the behaviour of the total current $J_r$ is explored in Sec. \ref{sec:J_tot}. We conclude with some open questions in Sec.~\ref{sec:conc}.

\section{Model and Results}\label{sec:model}

The symmetric exclusion processes (SEP) is a paradigmatic model for interacting particle systems \cite{Spitzer, Liggett} which have been used to describe a wide range of physical phenomena including particle transport in narrow channels, motion of molecular motors,  ion transport through porous medium etc. This process describes unbiased motion of particles on a lattice which interact via mutual local exclusion. In this section we define the dynamics of SEP with stochastic resetting and present a brief summary of our main results.

Let us consider a periodic lattice of size $L$ where each lattice site can contain at most one particle. The state of a site, say $x$, is characterized by a variable $s_x$ which takes values $1$ and $0$ depending on 
whether the site $x$ is occupied or not, respectively.   The configuration of the system is characterized by ${\cal C}=\{s_x; x=0,1,2,\dots L-1 \}.$ We consider the case of half-filling, \ie, 
the total number of particles $\sum_x s_x = \frac L2.$  The system evolves according to the following two dynamical moves:

\begin{itemize}
\item {\bf Hopping}: A particle randomly hops to one of its nearest neighbouring sites with unit rate, provided the target site is empty. 
\item {\bf Resetting}: In addition,  the system is `reset' to some specific configuration ${\cal C_0}$ with rate $r.$ In the following we consider ${\cal C_0}$ to be a step like state where all the particles are in the left-half of the lattice: 
\bea
{\cal C_0}:=\left \{ \begin{split}
                   s_x=1~& ~\text{for}~0 \le x \le \frac L2-1,\\
                   s_x=0 ~&~\text{otherwise.}  
                    \end{split}
 \right.
 \label{eq:C0}
\eea
\end{itemize}
Both the hopping and resetting dynamics conserve the total number of particles, so that the half-filling condition is respected at all times, and the global particle density remains fixed at $1/2.$  
 Between two resetting events the time evolution of the system is governed by the  hopping dynamics only. The time-scale associated with the resetting mechanism is given by $r^{-1},$ which also gives a measure of the typical duration between two consecutive resetting events.  Figure~\ref{fig:snapshot} shows typical examples of the time evolution for two different values of the resetting rate $r.$

In the absence of resetting, the master equation governing the time-evolution of the probability ${\cal P}_0({\cal C},t)$ for the system to be in the configuration ${\cal C}$ at time $t$ is given by 
\bea
\frac{\id }{\id t} {\cal P}_0({\cal C},t) = {\cal L}_0 {\cal P}_0({\cal C},t).
\eea
Here ${\cal L}_0$ is the Markov matrix in absence of the resetting, \ie, ${\cal L}_0 {\cal P}_0({\cal C},t) = \sum_{\cal C'}[W_\cal{C' \to C}{\cal P}_0({\cal C',t}) - W_\cal{C \to C'} {\cal P}_0({\cal C},t)]$ where $W_\cal{C' \to C}$ denotes the rate for the jump $\cal {C \to C'}$ due to hopping dynamics {\it only}. Note that, $W_\cal{C' \to C}=1$ only if the two configurations $\cal C$ and $\cal {C'}$ are connected by a single hop of a particle to a neighbouring site.

Let ${\cal P}({\cal C},t)$ denote the probability of finding the system in the configuration ${\cal C}$  at time $t$ in presence of resetting. In this case, the master equation reads,
\bea
\frac{\id }{\id t} {\cal P}({\cal C},t) &=& {\cal L}_0 {\cal P}({\cal C},t) \cr
&&+ r \sum_{\cal{C' \ne C}_0} {\cal P}({\cal C'},t) \delta_{\cal {C,C}_0} - r {\cal P}({\cal C},t) (1-\delta_{\cal{C,C}_0}) \cr
&=& ({\cal L}_0-r) {\cal P}({\cal C},t) + r \delta_{C,C_0}, \label{eq:Master_r}
\eea
where $\delta_{\cal{C,C}_0}$ is the Kronecker delta symbol, which takes the value unity when $\cal C$ is same as $\cal C_0,$ and is zero otherwise.
It is straightforward to write a formal solution of Eq.~\eqref{eq:Master_r},
\bea
 {\cal P}({\cal C},t) &=& e^{(\cal L_0-r)t} {\cal P}(\cal C,0) + r \int_0^t \id s ~e^{(\cal L_0-r)s} \delta_{C,C_0}\cr
 &=& e^{-rt} {\cal P}_0(\cal C,t) + r \int_0^t \id s ~e^{-rs} {\cal P}_0(\cal C,s). \label{eq:renewal_PC}
\eea
Here ${\cal P}_0(\cal C,t)=e^{\cal L_0 t}{\cal P}(\cal C,0)$ is the probability of finding the system in configuration $\cal{C}$ at time $t$ in the {\it absence of resetting} given that the system was initially at $\cal{C}_0,$ \ie, $\cal P(\cal C,0)={\cal P}_0(\cal C,0)=\delta_{\cal C, \cal C_0}$. Equation~\eqref{eq:renewal_PC} is nothing but the renewal equation for the configuration probability, which has been obtained earlier and used to study resetting phenomena in various other contexts \cite{Brownian,kpz-reset}. Note that Eq.~\eqref{eq:renewal_PC} holds true irrespective of the specific choice of $\cal C_0$ given in Eq.~\eqref{eq:C0}.

% 
% Let $W_\cal{C' \to C}$ denote the rate for the jump $\cal {C \to C'}$ due to hoppong dynamics {\it only}, \ie, $W_\cal{C' \to C}=1$ only if the two configurations $\cal C$ and $\cal {C'}$ are connected by a single hop of a particle to a neighbouring site. Then, the probability $P({\cal C},t),$ for system to be in the configuration ${\cal C}$ a time $t$ evolves following the master equation, 
% \bea
% \frac{\id }{\id t} P({\cal C},t) &=& \sum_{\cal C'}[W_{\cal {C' \to C}}P({\cal C',t}) - W_\cal{C \to C'} P({\cal C},t)] + r \sum_{\cal{C' \ne C}_0} P({\cal C'},t) \delta_{\cal {C,C}_0} - r P({\cal C},t)(1-\delta_{\cal{C,C}_0})\cr
% &=& \sum_{\cal C'}[W_\cal{C' \to C}P({\cal C',t}) - W_\cal{C \to C'} P({\cal C},t)] + r \delta_{C,C_0} - r P({\cal C},t)
% \eea

In the absence of resetting the ordinary SEP on a ring relaxes to an equilibrium state with flat density profile and zero current. The approach to the equilibrium state, starting from the step-like initial configuration ${\cal C}_0,$ is characterized by a diffusive current flowing through the system. It has been shown that, for an infinitely large system, the time-integrated current measuring the net particle flux through the central bond up to time $t,$ grows as $\sqrt{t}$ for large $t$ \cite{Derrida1,Derrida2}. Presence of resetting is expected to affect these characteristics of SEP which we investigate in detail in this paper. A brief summary of our results is presented below. 

\begin{itemize}
\item First,  we  compute an exact expression for the evolution of the average density profile $\rho(x,t) = \la s_x(t) \ra$ for any arbitrary value of the resetting rate $r,$ which is given in Eq.~\eqref{eq:rhor_xt}. 
We observe that the evolution is non-trivially modified due to the presence of resetting which leads to an inhomogeneous stationary density profile [see Fig.~\ref{fig:rho}(b)] in contrast to 
the flat one for $r=0$. 

\item This inhomogeneous density profile provides some characterization of the non-equilibrium state of the system. It is, however, also important to look at how the particle currents in the system are affected by the introduction of resetting. In addition to the usual diffusive current $J_\id(t)$ created due to the local hopping of the particles, there is also a contribution $\Jres(t)$ to the total current due to the global movements of the particles during the resetting events.

We show that the behaviour of the diffusive current changes drastically in presence of resetting. In particular, we compute the average diffusive current $\la J_\text{d}(t) \ra$ exactly, which, in the long time limit, shows a linear growth with time $t,$  
$$\la J_\text{d}(t) \ra \simeq t\sqrt{\frac r{r+4}}.$$  This behaviour is in stark contrast to the $\sqrt{t}$ growth which is seen in absence of resetting \cite{Derrida1}. Similar change in the dynamical behaviour  is also observed for the variance of the diffusive current,  which also grows as $\sim t$ in presence of resetting,  as opposed to $\sqrt{t}.$  We explore the behaviour of the resetting current $\Jres$ too and show that, its average and variance also grow linearly with time. 
We also investigate the probability distribution of $\Jd(t)$ and demonstrate that, in the long-time regime, the typical fluctuations of $J_\text{d}(t)$ around its mean is characterized by a Gaussian distribution. Similar Gaussian fluctuations are also expected for the resetting current $\Jres.$

%This inhomogeneous density profile is associated to a nonzero diffusive current $J_d(t)$ (created due to the local hopping of the particle).  However, we observe that introduction of resetting mechanism into the dynamics creates a drastic change in the growth behavior that $\la J_d(t)\ra$ grows linearly as $\la J_\text{d}(t) \ra \simeq t\sqrt{r/(r+4)}$. In it known that in absence of resetting this current on an average grows as $\sqrt{t}$ \cite{}.

%Using the above result, we calculate the instantaneous diffusive current flowing through the system, which in turn yields the average time-integrated current $\la J_\text{d}(t) \ra$. We show that, in the long time limit, $\la J_\text{d}(t) \ra \simeq t\sqrt{r/(r+4)}$ --- presence of resetting results in a linear increase of the diffusive current with time (in drastic contrast to the $\sqrt{t}$ behaviour for $r=0$).  We also calculate the variance of the diffusive current for small $r,$ and show that it also increases linearly with time $t.$ 

\item Finally, we study the behaviour of the total current $J_r = \Jd + \Jres$ and calculate the average $\la J_r(t)\ra$ and the second moment $\la J_r^2(t)\ra$ as functions of time $t.$  In the long-time limit the moments reach stationary values. In particular, we show that the average stationary current is given by
\bea
\la J_r \ra = \frac 1{\sqrt{r(r+4)}}. 
\eea
We also compute the stationary probability distribution of the total current $P_r^\text{st}(J_r),$ for small values of $r,$ using a renewal approach. Interestingly, it turns out that, this distribution is non-Gaussian, and has very asymmetric behaviour at the two tails.

% We also show that, in the long time limit, the distribution of $J_r$ becomes independent of time. We calculate this stationary distribution $P(J_r)$ for small values of $r$ analytically  which turns out  to be non-Gaussian. Interestingly, we find that,  this distribution describing  the typical fluctuations has a scaling form,
% \bea
%  P(J_r) = \frac 1{\sigma_r} {\cal P}\bigg(\frac {J_r - \mu_r}{\sigma_r}\bigg)
% \eea
% for typical values of $J_r$ around the stationary mean $\mu_r;$ here  $\sigma_r^2$ is the stationary values of the variance of $J_r.$ 

\end{itemize}

\section{Density profile}\label{sec:density}

Presence of repeated resetting to the inhomogeneous configuration $\cal C_0$ destroys the translational invariance in the system and a non-trivial density profile can be expected, even in the stationary state. 
%In this section we calculate the stationary density profile, along with the approach
%
The average density $\rho(x,t) = \la s_x(t) \ra$ is given by the probability that the site $x$ is occupied any time $t.$ The time-evolution equation for the density profile can be derived by multiplying Eq.~\eqref{eq:Master_r} by $s_x$ and summing over all configurations $\cal C,$
\bea
\frac{\id}{\id t} \rho(x,t) &=& \rho(x+1,t)+ \rho(x-1,t)- 2\rho(x,t) \cr
&& -r\rho(x,t) + r  \phi(x) \label{eq:drho}
\eea
Here $\phi(x)$ is the  density profile corresponding to the resetting configuration ${\cal C}_0$ which, as mentioned before, is also taken as the initial profile.  
The exact time-dependent density profile $\rho(x,t)$ can be obtained by solving Eq.~\eqref{eq:drho}. To this end we introduce the discrete Fourier transform
\bea
\tilde \rho(n,t) = \sum_{x=0}^{L-1} e^{i \frac{ 2\pi n x}{L}} \rho(x,t), \; \text{with}~ n=0,1,2 \dots L-1. ~~ \label{eq:rho_dft}
\eea
Substituting Eq.~\eqref{eq:rho_dft} in Eq.~\eqref{eq:drho}, we get,
\bea
\frac{\id }{\id t}\tilde \rho(n,t) = - (\lambda_n+r) \tilde \rho(n,t) + r \tilde \phi(n) %\quad \text{with} \quad \lambda_n= 2\left(1- \cos \frac{2 \pi n}{L}\right) 
\label{eq:dtilde_rho}
\eea
with $\lambda_n= 2\left(1- \cos \frac{2 \pi n}{L}\right)$ and $\tilde \phi(n)$ is the Fourier transform of the resetting (and initial) profile $\phi(x).$ Equation~\eqref{eq:dtilde_rho} can immediately be solved,
\bea
\tilde \rho(n,t) =  \frac{r\tilde \phi(n)}{r+\lambda_n } + \frac{\lambda_n \tilde \phi(n)}{r+\lambda_n} e^{-(r+\lambda_n)t}
\eea
The density profile is then obtained by inverting the Fourier transform,
\bea
\rho(x,t) &=& \frac rL \sum_{n=0}^{L-1} \frac{\tilde \phi(n)}{r+\lambda_n } e^{-i \frac{2 \pi n x}L} \cr
&&+ \frac 1L \sum_{n=0}^{L-1}\frac{\lambda_n \tilde \phi(n)}{r+\lambda_n} e^{-(r+\lambda_n)t} e^{-i \frac{2 \pi n x}L}\label{eq:rhor_xt}
\eea

\begin{figure}
 \centering
  \includegraphics[width=8.8 cm]{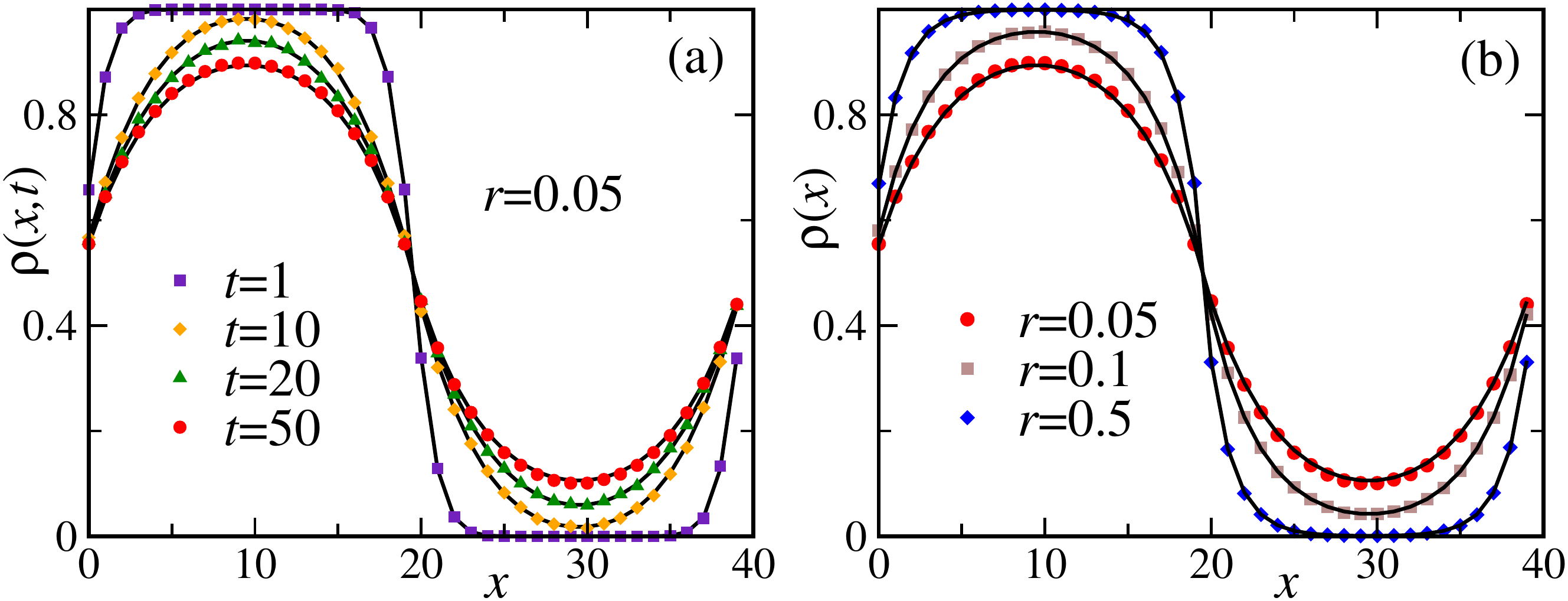}
 % reset_profile.pdf: 0x0 px, 300dpi, 0.00x0.00 cm, bb=
 \caption{Density profile:  (a) Time evolution of the density profile $\rho(x,t),$ starting from a step-like initial condition, for resetting rate $r=0.05,$ and for different values of time $t.$  The symbols correspond to the data obtained from numerical simulations and the solid lines correspond to the analytical result [see Eq.~\eqref{eq:rhor_t_C0}]. (b) The stationary density profile $\rho(x)$ for different reset rates $r.$ The symbols correspond to the data obtained from numerical simulations and the solid lines correspond to the analytical result [see Eq.~\eqref{eq:rhor_stat_C0}]. The lattice size $L=40$ for both (a) and (b).}
 \label{fig:rho}
\end{figure}

In the stationary state, the second term decays exponentially and the stationary density profile is given by,
\bea
\rho(x) &=& \frac rL \sum_{n=0}^{L-1} \frac{\tilde \phi(n)}{r+\lambda_n } e^{-i \frac{2 \pi n x}L} \cr
&=& \frac 12 + \frac rL \sum_{n=1}^{L-1} \frac{\tilde \phi(n)}{r+\lambda_n } e^{-i \frac{2 \pi n x}L} \label{eq:rhor_stat}
\eea
where we have used the fact that $\tilde \phi(0) = \sum_x \phi(x) = L/2.$ Clearly, in absence of resetting, \ie for $r=0,$ we get the flat profile which corresponds to the equilibrium scenario. For non-zero $r,$ however, the stationary profile is non-trivial and corresponds to a non-equilibrium stationary state, carrying non-zero current.  We will explore that in the next section.

It is worth mentioning that $\rho(x,t)$ also satisfies a renewal equation (following directly from Eq.~\eqref{eq:renewal_PC}) in terms of the density profile $\rho_0(x,t)$ in absence of resetting,
\bea
\rho(x,t) = e^{-rt} \rho_0(x,t) + r \int_0^t \id \tau ~e^{- r \tau} \rho_0(x,\tau). \label{eq:rhoreset}
\eea
For the sake of completeness we have added a brief review of the density profile and its evolution for ordinary SEP in the Appendix \ref{app:sep}. Using the explicit form of $\rho_0(x,t)$ given in Eq.~\eqref{eq:rho0xt} it is straightforward to check that Eq.~\eqref{eq:rhoreset} leads to Eq.~\eqref{eq:rhor_xt}. 

We have not used any specific form of $\phi(x)$ so far; in fact, the results above are valid for resetting to any generic profile. For the specific choice of the step-like configuration given in Eq.~\eqref{eq:C0} we have $\phi(x)=1- \Theta(x+ 1 -\frac L2)$ and
\bea
\tilde \phi(n) = \frac{1-(-1)^n}{1-e^{i \frac{2 \pi n}L}} = \left\{
\begin{split}
1+i ~\cot \frac{\pi n}L & \qquad \text{for odd} ~~~ n \cr
0~~~~~ &\qquad \text{for even} ~~~ n                                                        
\end{split}
\right.
\eea
% \bea
% \tilde \phi(n) = \frac{1-(-1)^n}{1-e^{i \frac{2 \pi n}L}} = \left\{
% \begin{split}
% \frac2{1-e^{\frac{i 2 \pi n}{L}}} & \qquad \text{for odd} ~~~ n \cr
% 0~~~~~ &\qquad \text{for even} ~~~ n                                                        
% \end{split}
% \right.
% \eea
In that case, the density profile takes the form,
\bea
\rho(x,t) &=& \rho(x) + \frac 1L \sum_{n=1,3}^{L-1} e^{- i  \frac{2 \pi n x}{L}} \frac{\lambda_n (1+i ~\cot \frac{\pi n}L) }{r+\lambda_n}  e^{-(r+\lambda_n)t}\cr
&& \label{eq:rhor_t_C0}
\eea
where 
\bea
\rho(x) = \frac 12 + \frac rL \sum_{n=1,3}^{L-1} e^{- i  \frac{2 \pi n x}{L}} \frac {(1+i ~\text{cot} \frac{\pi n}L)}{r+\lambda_n} \label{eq:rhor_stat_C0}
\eea
is the stationary profile.

Figure~\ref{fig:rho}(a) shows the time-evolution of the density profile $\rho(x,t)$ for a specific resetting rate $r$ and Fig.~\ref{fig:rho}(b) shows stationary profiles $\rho(x)$ for different values of $r.$ In both cases, the analytical results (solid lines) are compared with the data obtained from numerical simulations (symbols). An excellent match confirms our analytical prediction. 

\section{Particle current}\label{sec:cur}

The behaviour of current plays an important role in characterizing the interacting particle systems like exclusion processes. 
For ordinary SEP, there is no particle current flowing through the system in the stationary (equilibrium) state. However, starting from a step-like initial configuration, the relaxation to equilibrium is characterized by the presence  of a non-vanishing particle current. In particular, the behaviour of the time-integrated current, \ie, the net particle flux through the central bond up to time $t,$ has been studied extensively in the past and it was shown that at long time limit, the average flux grows $\sim \sqrt{t}$ \cite{Derrida1,Derrida2}.

%In this section we investigate the behaviour of particle current in SEP when resetting dynamics is added.

%We are particularly interested in the particle flux passing thorugh the central bond $(\frac L2 -1, \frac L2)$ over the time duration $[0,t].$

\begin{figure}[t]
 \centering
 \includegraphics[width=8 cm]{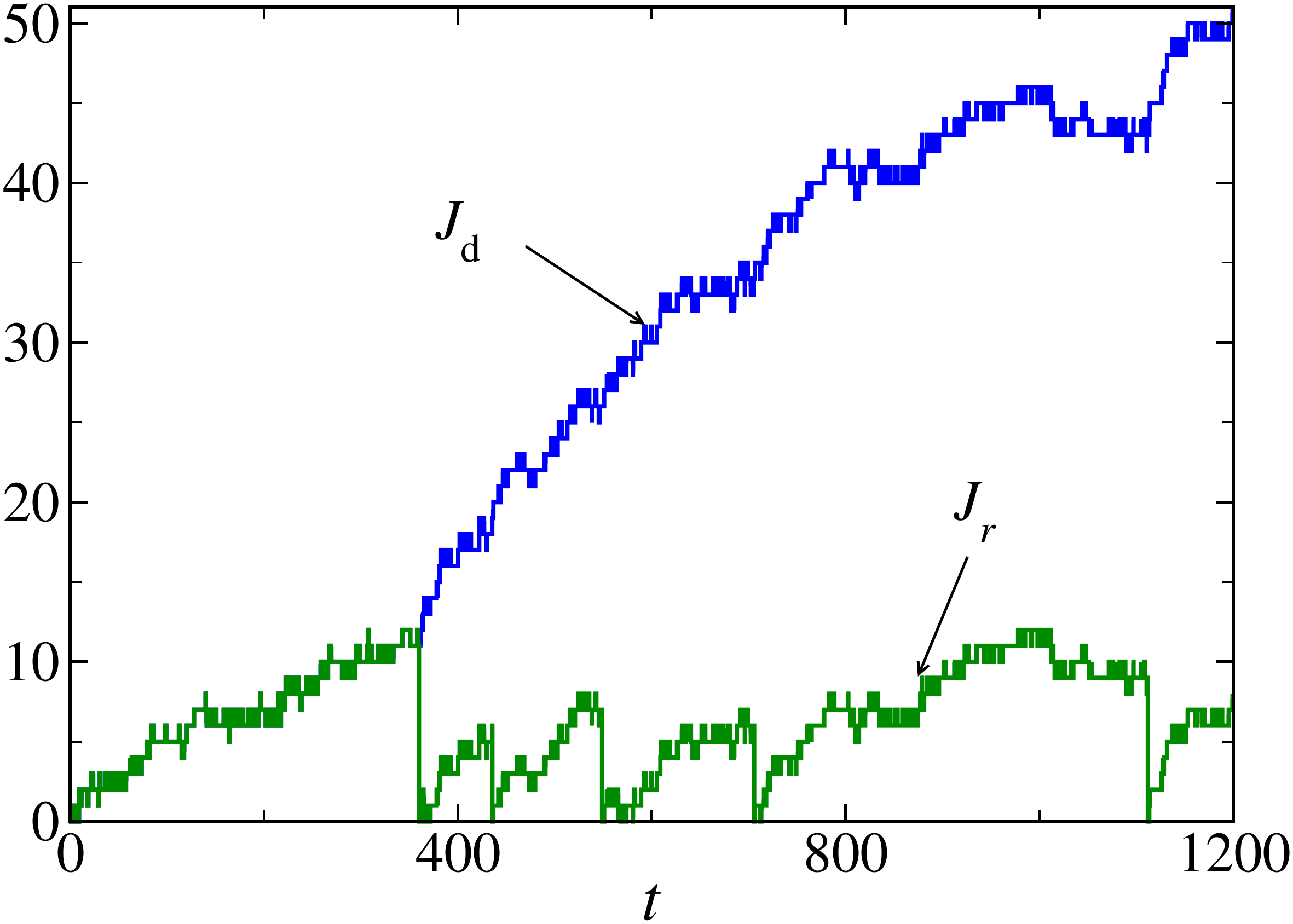}
 % traj.pdf: 0x0 pixel, 300dpi, 0.00x0.00 cm, bb=
 \caption{Time evolution of the  diffusive current $\Jd(t)$   and the total current $J_r(t)$ along a typical trajectory of the system. On an average, the diffusive current increases with time $t.$ The total current vanishes after each resetting event --indicated by the vertical lines on the light green curve -- and reaches a stationary state in the long time limit. }
 \label{fig:traj}
\end{figure}

In presence of resetting, there are two different kinds of particle motions, consequently the total current can be expressed as,
\bea
J_r(t) = J_\text{d}(t) + \Jres(t) \label{eq:Jtot_def}
\eea
Here $\Jd$ is net diffusive flux, \ie, the net number of particles which crossed the central bond due to the nearest neighbour hopping. $\Jres$ denotes the contribution due to the sudden
reset to the step-like configuration $\cal C_0$. Note that, after a resetting, the system is brought back to $\cal C_0$, \ie, there are no particles to the right of the central bond, implying that the total current $J_r$ is also reset to zero after each resetting event. Figure~\ref{fig:traj} shows the time evolution of $J_\text{d}$ and $J_r$ for a typical trajectory of the system. The sudden jumps in $J_r$ indicate the resetting events. 

%Between two resetting events the system evolves as an ordinary 

% Note that, during a reset, the particles which had crossed the central bond due to hopping since the previous resetting...

In the absence of resetting, the only source of current is the diffusive hopping motion. In the following we explore the behaviours of all these three different currents, in presence of resetting.

\begin{figure*}[t]
 \centering
  \includegraphics[width=11.4 cm]{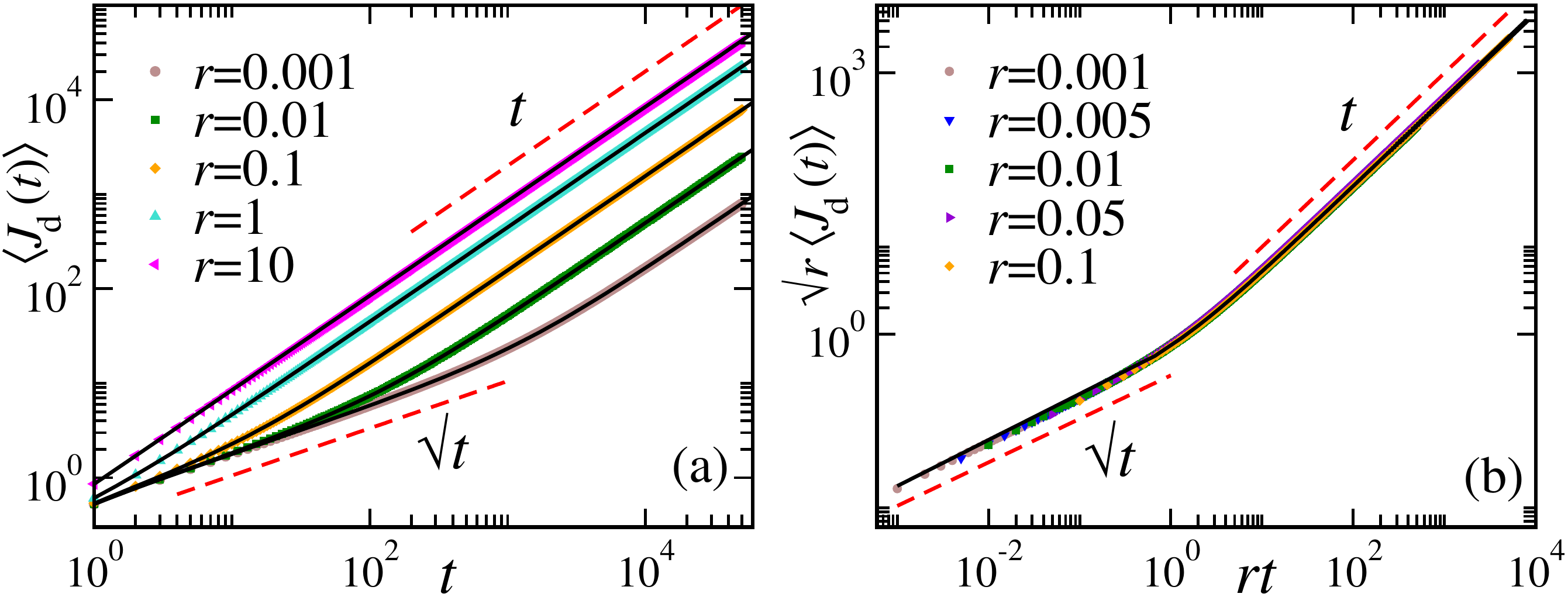} \includegraphics[width=5.6 cm]{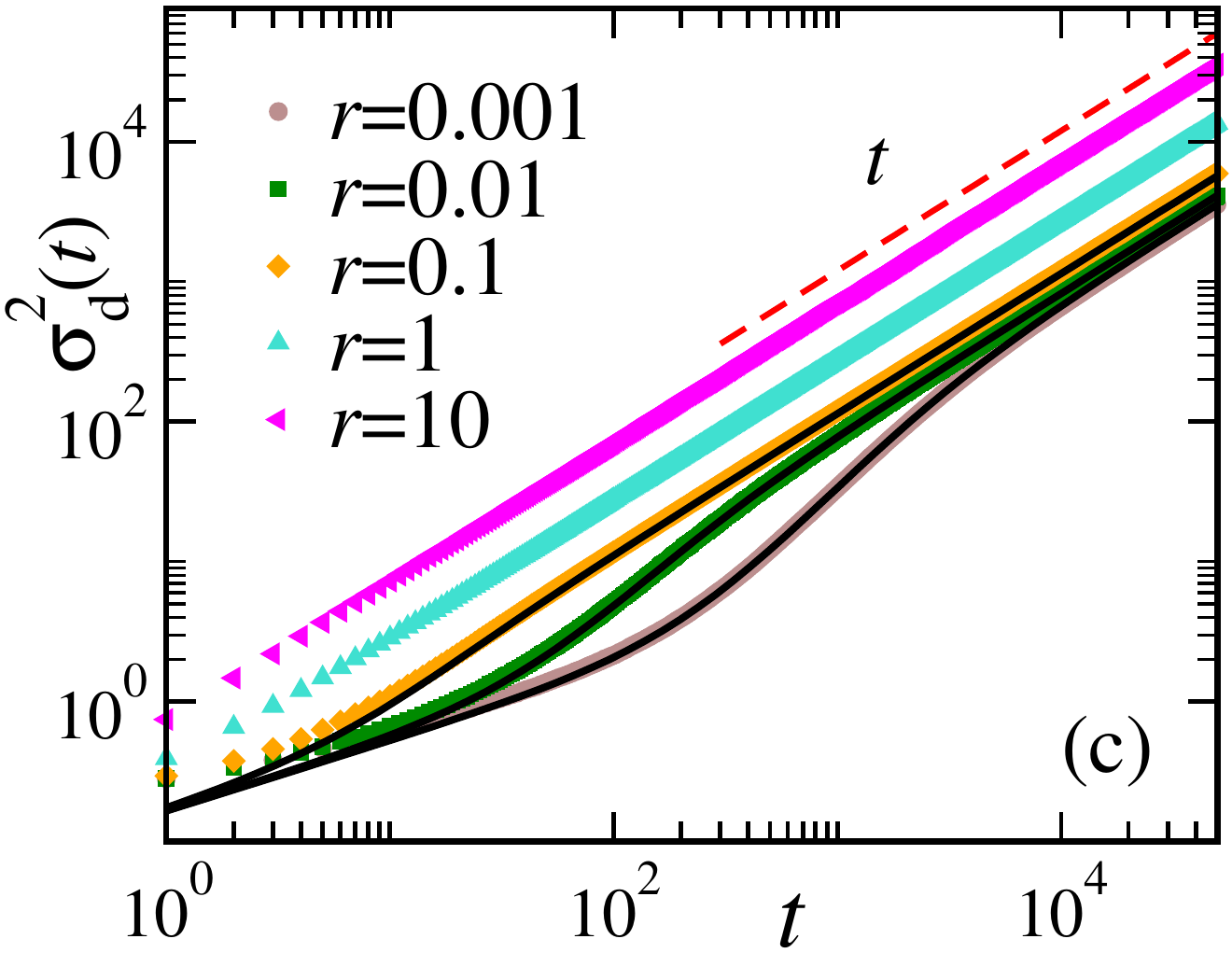}
 % reset_profile.pdf: 0x0 px, 300dpi, 0.00x0.00 cm, bb=
 \caption{Behaviour of the diffusive current $\Jd.$ (a) Plot of $\la J_\text{d}(t)\ra$ as a function of time $t$ for different values of $r.$  The symbols correspond to the data obtained from numerical simulations whereas the solid lines correspond to the analytical prediction [see Eq.~\eqref{eq:avJdif_t}]. The lowermost curve corresponds to the smallest value of $r.$ 
 (b) Scaling collapse of $\sqrt{r}\la \Jd(t)\ra$ for small $r$ according to Eq.~ \eqref{eq:Jd_av}; the solid line corresponds to the predicted scaling function.  (c) Plot of the variance $\sigma_\id^2(t)$ {\it vs} $t$ for different values of $r.$ The curves corresponding to small values of $r$ (three lower curves) are compared with the analytical predictions (solid lines). A lattice of size $L=1000$ is used here for the numerical simulations.}
 \label{fig:J_r}
\end{figure*}

\subsection{Diffusive current}\label{sec:J_dif}

The diffusive current $\Jd(t)$ measures the total number of particles which crossed the central bond $(\frac L2 -1, \frac L2)$ during the time interval $[0,t],$ and can be expressed as a time integral,
\bea
J_\text{d}(t) &=& \int_0^t \id s~ j(s) \label{eq:Jdif_intdefn}
\eea
Here, $j(t)$ denotes the instantaneous diffusive current, \ie, the number of particles crossing the central bond during the time-interval $t$ and $t+\id t.$ The average instantaneous current is given by,
% 
% The mean instanteneous diffusive current through the central bond  
% $(\frac L2 -1, \frac L2)$ is given by the average particle flux across the bond during the time-interval $t$ and $t+\id t,$
\bea
\la j(t) \ra &= & \left\la s_{\frac L2-1}(1-s_{\frac L2})\right\ra - \left\la (1- s_{\frac L2-1})s_{\frac L2}\right\ra \cr
&= &\rho\left(\frac L2 -1,t\right) - \rho\left(\frac L2,t \right)
\eea
Using the explicit expression for the density from eq.~\eqref{eq:rhor_t_C0} one gets,
\bea
\la j(t) \ra = \frac 2L \sum_{n=1,3}^{L-1}\bigg[\frac r {r+ \lambda_n}+ \frac {\lambda_n}{r+\lambda_n}e^{-(r+\lambda_n)t}\bigg].
\eea
In the limit of thermodynamically large system size, \ie $L \to \infty$, the sum in the above expression can be converted to an integral over continuous variable $q= 2 \pi n/L$  and we get, 
\bea
\la j(t) \ra = \int_0^{2 \pi} \frac {\id q}{2 \pi} \bigg[\frac r{r+\lambda_q}+ \frac {\lambda_q}{r+\lambda_q}e^{-(r+\lambda_q)t}\bigg]
\eea
where $\lambda_q = 2(1- \cos q).$ In the long-time regime, the second term decays exponentially and $\la j(t)\ra$ reaches a stationary value,
\bea
\lim_{t\to \infty} \la j(t) \ra = \int_0^{2 \pi} \frac {\id q}{2 \pi} \frac r{r+ 2(1- \cos q)} = \sqrt{\frac r{r+4}}. \label{eq:jr_stat}
\eea
The average net flux $\la J_\text{d}(t)\ra$ up to time $t$ can be found by integrating the instantaneous current,
\bea
\la J_\text{d}(t)\ra &=& \sqrt{\frac r{r+4}} t \cr
& +& \int_0^{2 \pi} \frac {\id q}{2 \pi}\frac {\lambda_q}{(r+\lambda_q)^2}(1- e^{-(r+\lambda_q)t}). \label{eq:avJdif_q}
\eea
Clearly, in the long-time regime, the second term goes to a constant and the first term dominates the behaviour of the average current which grows linearly with time,
\bea
\la J_\text{d}(t)\ra \simeq \sqrt{\frac r{r+4}} t.
\eea
This equation is one of our main results, which shows that the behaviour of the diffusive current changes drastically by the presence of resetting; instead of the standard $\sqrt{t}$ growth in a diffusive system, resetting yields a much faster, linear, temporal growth of the  diffusive current.
The average current $\la \Jd(t)\ra$ at any time $t,$ \ie, before reaching the $\sim t$ behaviour, can be obtained from Eq.~\eqref{eq:avJdif_q} by evaluating the $q$-integral numerically. In fact, one can also derive an alternative expression which lends itself more easily to numerical evaluation. Let us recall that the density profile $\rho(x,t)$ satisfies a renewal equation \eqref{eq:rhoreset} for any $x$. Then, clearly, $\la j(t)\ra$ must also satisfy the same renewal equation,
\bea
\la j(t)\ra &=& e^{-r t} \la j_0(t) \ra + r \int_0^t \id \tau~ e^{-r \tau} \la j_0(\tau) \ra \label{eq:jrt_defn}
\eea
where $j_0(t)$ denotes instantaneous current through the central bond in the absence of resetting. The average instantaneous current is given by $\la j_0(t) \ra = e^{-2t} I_0(2t)$ where $I_0$ is the Modified Bessel function of the first kind (see Appendix \ref{sec:J0} for the details). The average diffusive net current is obtained by integrating the above equation w.r.t. time [see Eq.~\eqref{eq:Jdif_intdefn}],
% \bea
% \la \Jd(t) \ra &=& (1+rt)\int_0^t \id \tau ~e^{-r \tau} \la j_0(\tau) \ra \cr
% && -  r \int_0^t \id \tau ~\tau e^{-r \tau} \la j_0(\tau) \ra
% \eea
\bea
\la \Jd(t) \ra &=& \int_0^t \id \tau ~e^{-r \tau} (1+rt - r\tau) \la j_0(\tau) \ra. \;\; \label{eq:avJdif_t} 
\eea
It is straightforward to show that Eq.~\eqref{eq:avJdif_t} is equivalent to Eq.~\eqref{eq:avJdif_q}. Average current $\la \Jd(t) \ra$ computed from Eq.~\eqref{eq:avJdif_t}, for different values of $r,$ are plotted together with the same obtained from simulation in Figure \ref{fig:J_r}(a). 

% Figure \ref{fig:J_r}(a) shows comparison of $\la J_\text{d}(t)\ra$ computed from Eq.~\eqref{eq:avJdif_t}  for different values of $r,$ with those obtained from numerical simulations. A perfect match confirms the analytical prediction.

An explicit form for $\la J_\text{d}(t)\ra$ can be derived for small $r \ll 1.$ 
%In this case, the integral in Eq.~\eqref{eq:avJdif_t} is dominated by the contribution from large $\tau$ value. 
Using a variable transformation $w=r \tau,$ and using the exact form for $\la j_0(\tau) \ra$ we get,
\bea
\la \Jd(t) \ra &=& \frac 1r \int_0^{rt} \id w   (1 + rt - w)  e^{-w} e^{-\frac{2 w}r} I_0\left(\frac{2 w}r\right).\;\;\;
\eea
For small $r,$ the argument of $I_0$ is large and one can use the asymptotic form for the Modified Bessel function given in Eq.~ \eqref{eq:I01_large},
\bea
\la \Jd (t) \ra & \simeq & \frac 1r \int_0^{rt} \id w   (1 + rt - w)  e^{-w} \frac 1{2 \sqrt{\pi w/r}} \cr
&=&\frac 1{2 \sqrt{r}} \bigg[\left(r t+ \frac 12 \right) \text{erf}(\sqrt{rt})+ \sqrt{\frac{rt}{\pi}}e^{-rt}\bigg]. \label{eq:Jd_av}
\eea
In the short time-regime this function grows as $\sqrt{t}$ which is reminiscent of the free SEP and crosses over to the linear behaviour for $t \gg r^{-1}.$ Figure \ref{fig:J_r}(b) shows plot of $\sqrt{r}\la \Jd(t) \ra$ as a function of $r t$ for different small values of $r$ which shows a perfect collapse and matches with the scaling function given by the above equation. 

To characterize the fluctuation of the diffusive current we next calculate the second moment of $J_\text{d}.$ The above renewal equation method cannot be applied directly to compute higher order moments. To this end, we now adopt a different approach.  Let us assume that there are $n$ resetting events during the time interval $[0,t];$ moreover, let $t_i$ denote the interval between $(i-1)^{th}$ and $i^{th}$ events, so that $\sum_{i=1}^{n+1} t_i = t.$ Note that, $t_{n+1}$ denotes the time-interval between the last reset and the final time $t.$ Let us also recall that between two consecutive resetting events the system evolves following ordinary SEP dynamics.
The diffusive current during the interval $[0,t]$ can, then, be expressed as,
\bea
J_\text{d} = \sum_{i=1}^{n+1} J_0(t_i), \label{J_d-sum}
\eea
where, $J_0(t_i)$ are independent of each other. For notational convenience, we denote $J_i \equiv J_0(t_i).$ The probability density that the diffusive current will have a value $\Jd$ in time $t$ is then given by
\bea
P(J_\text{d},t) &=& \sum_{n=0}^{\infty} \int_{0}^t \prod_{i=1}^{n+1}\id t_i  ~ {\cal P}_n(\{t_i \};t)  \cr 
&& \times \int \prod_{i=1}^{n+1}\id J_i P_0(J_i,t_i) 
  ~\delta(J_\text{d}-\sum_i J_i), \quad \text{where,} \cr
%\text{where,} && \cr
{\cal P}_n(\{t_i \};t) &=& r^n e^{-r\sum_{i=1}^{n+1} t_i} \delta(t-\sum_i t_i) \label{eq:P_Jd}
\eea
denotes the probability of having $n$ resetting events with duration $t_i$ within the interval $[0,t].$ The distribution of the individual $J_i$-s are denoted by $P_0(J_i,t_i)$ which is exactly the distribution of the diffusive current in SEP, in the absence of resetting.

% It is then convenient to calculate the moment generating function,
% \bea
% \la e^{\lambda J_\text{d}} \ra &=& \sum_{n=0}^{\infty} \int_{0}^t \prod_{i=1}^{n+1} \id t_i e^{\lambda J_\text{d}} {\cal P}_n(\{t_i \};t) \prod_{i=1}^{n+1} P_0(J_i,t_i)\cr
% \eea

To handle the constraints presented by the $\delta$-functions, it is convenient to calculate the Laplace transform w.r.t. time $t$ of the moment generating function $\la e^{\lambda J_\text{d}} \ra,$
\bea
Q(s,\lambda) &=& {\cal L}_{t \to s} [\la e^{\lambda J_\text{d}} \ra]= \int_0^{\infty} \id t ~e^{-s t}\la e^{\lambda J_\text{d}} \ra \cr
&=& \int_0^{\infty} \id t ~e^{-s t}  \int \id J_\text{d} ~ e^{\lambda J_\text{d}} P(J_\text{d},t).
\eea
Using Eq.~\eqref{eq:P_Jd}, and performing the integrals over $\Jd$
and $t,$ we get,
\bea
Q(s,\lambda)&=& \sum_{n=0}^\infty r^n \int_0^\infty \prod_{i}^{n+1} \id t_i \exp {\bigg[-(r+s) \sum_{i=1}^{n+1} t_i\bigg]} \n \\ [0.2em]
&&\times \int \prod_{i}^{n+1} \id J_i \exp{\bigg[\lambda \sum_{i=1}^{n+1} J_i\bigg]} P_0(J_i,t_i) \cr
&=& \sum_{n=0}^\infty r^n h(s,\lambda)^{n+1} \label{eq:Q_sum}
\eea
where, we have denoted,
\bea
h(s,\lambda) = \int_0^\infty \id \tau ~e^{-(r+s)\tau} \int \id J_0 ~ e^{\lambda J_0} P_0(J_0,\tau)\label{eq:hsl} 
\eea
Performing the sum in Eq.~\eqref{eq:Q_sum}, we get,
\bea
Q(s,\lambda) = \frac{h(s,\lambda)}{1- r h(s,\lambda)} \label{eq:Qsl}
\eea
which gives a simple relation between the moment generating functions of the current in the presence and absence of resetting. 
To calculate $h(s,\lambda)$ we need the current distribution $P_0(J_0,\tau)$ for the ordinary SEP, which is not known in general for arbitrary values of $\tau$. However, for small values of $r$ and $s,$ the $\tau$-integral in Eq.~\eqref{eq:hsl} is dominated by large values of $\tau,$ and in that case one can use the result of Ref.~\cite{Derrida1} where the authors have derived an expression for the moment generating function of $J_0(\tau)$ in the large time limit.
Adapting their result to our specific case (see Appendix \ref{sec:J0_dist} for the details), we have,
\bea
\int \id J_0 ~ e^{\lambda J_0} P_0(J_0,\tau) =\la e^{\lambda J_0} \ra \simeq e^{\sqrt \tau F(\lambda)}, \label{eq:J0_gen}
\eea
with, 
\bea
F(\lambda) = -\frac 1{\sqrt \pi} \text{Li}_{3/2}(1-e^\lambda).\label{eq:F_lam}
\eea
Here $\text{Li}_\alpha(z)$ denotes the Poly-Logarithm function (see Ref.~\cite{polylog}, Eq.~25.12.10). Substituting Eq.~\eqref{eq:J0_gen} in Eq.~\eqref{eq:hsl} and performing the integral over $\tau,$ we get, for small $r$ and $s,$
\bea
h(s,\lambda) &=& \frac{1}{r+s}\bigg[1+ \frac{\sqrt \pi F(\lambda)}{2\sqrt{r+s}}e^{\frac{F(\lambda)^2}{4(r+s)}}\bigg(1+\text{erf}\bigg[\frac{F(\lambda)}{2\sqrt{r+s}}\bigg]\bigg) \bigg]\cr
&& \label{eq:hsl_F}
\eea
One can easily extract the Laplace transforms of the moments using Eq.~\eqref{eq:hsl_F} along with \eqref{eq:Qsl}. First, we have,
\bea
{\cal L}_{t \to s} \left[\la \Jd(t) \ra \right] &=& \frac{\id}{\id \lambda} Q(s,\lambda)\bigg|_{\lambda=0} \cr
&=& \frac{\sqrt{r+s}}{2s^2}
\eea
The average current can be obtained by inverting the Laplace transform,
\bea
\mu_\id(t) \equiv \la \Jd(t) \ra &=& {\cal L}^{-1}_{s \to t} \left[\frac{\sqrt{r+s}}{2s^2} \right]
\eea
The inversion can be performed exactly using Mathematica, and yields, 
\bea
\mu_\id(t)&=&  \frac 1{2 \sqrt{r}} \bigg[\left(r t+ \frac 12 \right) \text{erf}(\sqrt{rt})+ \sqrt{\frac{rt}{\pi}}e^{-rt}\bigg] \label{eq:Jd_av2} 
\eea
Note that the above equation is the same as Eq.~\eqref{eq:Jd_av}, which was obtained using a different method.

\begin{figure}[t]
 \centering
 \includegraphics[width=8.8 cm]{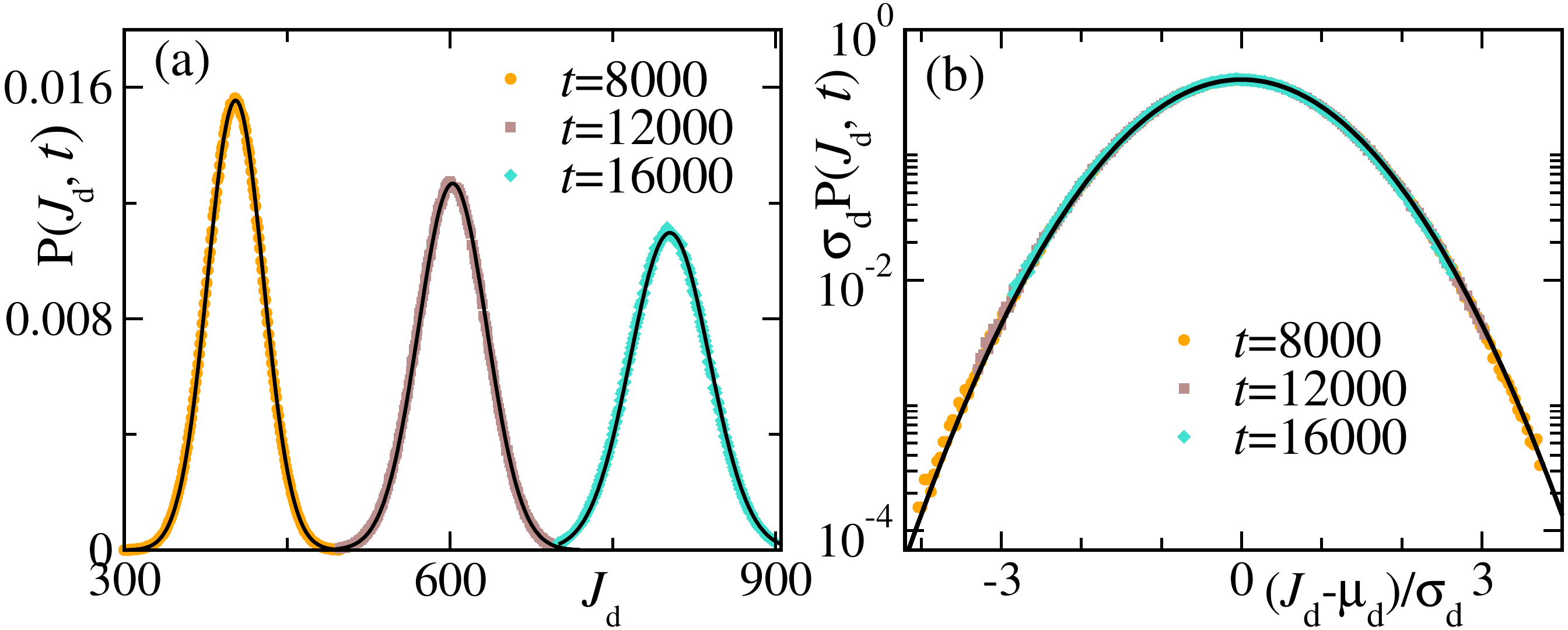}
 % dist_r_01.pdf: 0x0 px, 300dpi, 0.00x0.00 cm, bb=
 \caption{Distribution of the diffusive current $P(J_\id,t)$ for $r=0.01$: (a) Plot of $P(J_\id,t)$ {\it vs} $\Jd$ for different values of $t,$ the left most curve corresponding to the  smallest value of $t.$ 
 The solid lines correspond to the predicted Gaussian form \eqref{eq:PJd_gauss}.  (b) The same data are plotted as a function of $(J_\id - \mu_\id)/\sigma_d.$ The solid black line corresponds to a standard normal distribution ${\cal N}(0,1).$ }
 \label{fig:dist_r0.01}
\end{figure}

The Laplace transform of the second moment is obtained from the second derivative of $Q(s,\lambda),$
\bea
{\cal L}_{t \to s} \la \Jd^2(t) \ra &=& \frac{\id^2}{\id \lambda^2} Q(s,\lambda)\bigg|_{\lambda=0}\cr
%&=& \frac{b\pi s(r+s) + \sqrt{r+s}(\pi r+s)}{2 \pi s^3 \sqrt{r+s}}
&=& \frac 1{\pi s^2}+ \frac{b \sqrt{r+s}}{2s^2} + \frac r{2 s^3} 
\eea
where $b=(1-1/\sqrt{2}).$ Fortunately, the inverse Laplace transform can be performed exactly  in this case also, and it yields, for small $r,$ 
\bea
\la \Jd^2(t) \ra &=& \frac 1{4 \pi}\bigg[t(\pi r t+4)+2 b \sqrt{\pi t} e^{-r t}\cr
&& \quad + \frac{b \pi }{\sqrt{r}}(1+2 rt)\text{erf}(\sqrt{rt})\bigg] \label{eq:Jdsq_av}
\eea
Note that, the above expression is expected to be valid for large $t,$ as we have assumed $s$ to be small. The variance of the diffusive current $\sigma^2_\text{d}(t) =\la \Jd^2(t) \ra -\la \Jd(t) \ra^2$ can be obtained using Eqs.~\eqref{eq:Jd_av2} and \eqref{eq:Jdsq_av}. In particular, in the long time limit, the variance increases linearly with time $t$, and is given by,
\bea
\sigma_\text{d}^2(t) \simeq t \bigg[\frac{4-\pi}{4 \pi} + \frac {\sqrt r}2\bigg(1- \frac 1{\sqrt{2}}\bigg)\bigg]
\label{eq.sigma_J_d}
\eea
Figure \ref{fig:J_r}(c) shows a plot of $\sigma^2_\text{d}(t)$ vs $t$ for different values of $r,$ obtained from numerical simulations; all the curves show linear growth in the long time regime. The curves corresponding  to small values of $r \ll 1$ are compared with the analytical result (solid lines) which shows a perfect match for $t > 10.$ \\

\noindent {\bf Distribution of $\Jd$}: It is interesting to investigate the probability distribution of the diffusive current $J_d(t)$.  From Eq.~\eqref{J_d-sum} we observe that $J_d(t)$ is the sum of the hopping currents $J_0(t_i)$  between successive resetting events. Since the time-evolution of the system is Markovian and after each resetting event the system is brought back to the initial configuration, the variables $J_0(t_i)$ are independent and distributed identically \footnote{Of course, the time durations $\{ t_i\}$ are different.}. As mentioned earlier, the distribution of $J_0(t_i)$ is known \cite{Derrida1} and  has finite moments. 
Over a large time interval $t$, the number $n$ of resetting events, which is also a random quantity, is typically large  and on an average grows linearly with time $t;$ in fact, $\la n\ra = rt.$  For $t \gg r^{-1},$   $\Jd(t)$ is a sum of large number of independent random variables. Hence, by central limit theorem, one can expect that for large $t$, the typical distribution of $J_d$ would be a Gaussian: 
\bea
P(\Jd,t) \simeq \frac{1}{\sqrt{2 \pi \sigma_d^2(t)}}\exp\left(-\frac{[J_d-\mu_\id(t)]^2}{2 \sigma_d^2(t)}\right) \label{eq:PJd_gauss}
\eea
where the mean $\mu_\id(t)$ and the variance $\sigma_d^2(t)$ are given in Eqs.~\eqref{eq:Jd_av2} and \eqref{eq.sigma_J_d} respectively. This prediction is verified in Fig.~\ref{fig:dist_r0.01}(a) where the Gaussian form of $P(\Jd,t)$ is compared to the data obtained from numerical simulations for a set of (large) values of $t$ and fixed $r.$ Clearly, the analytical curves are indistinguishable from the simulation data, which confirms our prediction.   Fig.~\ref{fig:dist_r0.01}(b) shows the same data plotted against the scaled variable $(\Jd-\mu_d(t))/\sigma_d(t)$ and compared with the standard normal distribution (solid black line).

\subsection{Resetting current}\label{sec:J_reset}

Presence of the resetting dynamics gives rise to a resetting current $\Jres$ [see Eq.~\eqref{eq:Jtot_def}] which measures the flow of particles due to the sudden change in the configuration of the system. In this Section we investigate the properties of this resetting current. 
Let us remember that the number of particles crossing the central bond (from right to left) at the resetting event is exactly same as the hopping current (from left to right) during the period after the previous resetting event. Net resetting current during a time interval $[0,t],$ then, can be expressed as,
\bea
\Jres = - \sum_{i=1}^n{J_0(t_i)} \label{eq:Jreset_def}
\eea
where, as before, $n$ denotes the number of resetting events in time $t$ and $t_i$ denotes the interval between $(i-1)^{th}$ and $i^{th}$ resetting events. 
Note that, the upper limit of the sum is $n$ in the above  Eq.~\eqref{eq:Jreset_def} as there is no contribution to the resetting current after the last resetting event.

\begin{figure}
 \centering
 \includegraphics[width=8.8 cm]{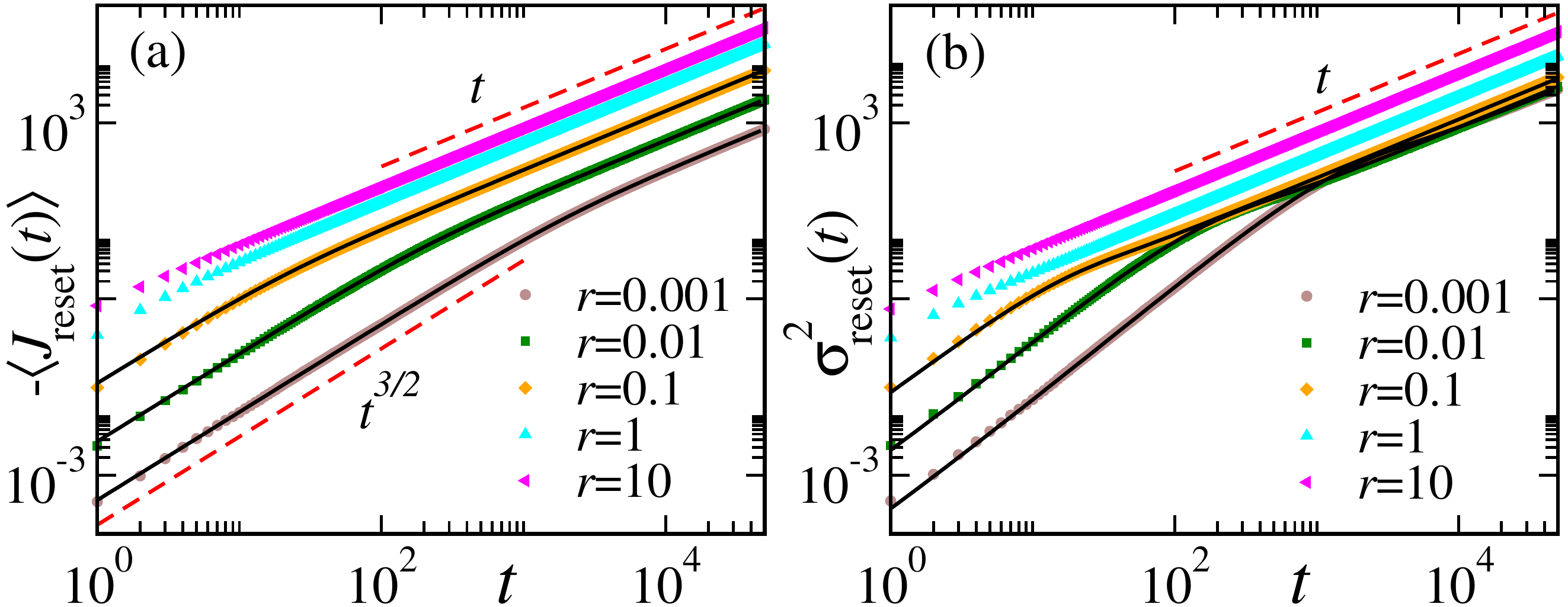}   
 \caption{Behaviour of the resetting current $\Jres :$ (a) Plot of  $-\la \Jres(t)\ra$ as a function of time for different values of $r$ obtained from numerical simulations. The lowermost curve corresponds to the smallest value of $r.$ The black solid lines correspond to the analytical prediction Eq.~\eqref{eq:Jreset_av} for small values of $r.$  (b) Variance of the resetting current $\la \sigma^2_\text{reset}(t) \ra$ as a function of time $t$ for different values of $r.$ The curves corresponding to small values of $r$ (three lower curves) are compared with the analytical result (black solid lines).  A lattice of size $L=1000$ is used for all the numerical simulations.}
 \label{fig:Jreset}
\end{figure}

To calculate the moments of $\Jres$ we follow the same method as in Sec.~\ref{sec:J_dif} and calculate the Laplace transform of the moment generating function of $\Jres,$
\bea
K(s,\lambda) = \int_0^{\infty} \id t ~e^{-s t} \int \id \Jres e^{\lambda \Jres} \mathscr P(\Jres,t).~~\label{eq:K-def}
\eea
Here $\mathscr P(\Jres,t)$ denotes the probability that the resetting current has a value $\Jres$ at time $t,$ and is given by,
\bea
\mathscr P(\Jres,t) &=& \sum_{n=0}^{\infty} \int_{0}^t \prod_{i=1}^{n+1}\id t_i  ~ {\cal P}_n(\{t_i \};t)  \cr 
& \times&  \int \prod_{i=1}^{n}\id J_i P_0(J_i,t_i) 
  ~\delta(\Jres+\sum_{i=1}^n J_i).~~ \label{eq:P_Jreset}
\eea
with ${\cal P}_n(\{t_i \};t)$ given in Eq.~\eqref{eq:P_Jd}. As before, we have used $J_i \equiv J_0(t_i).$ Using Eq.~\eqref{eq:P_Jreset} in Eq.~\eqref{eq:K-def} and performing the integrals over $t$ and $\Jres,$ we get, 
\bea
K(s,\lambda) &=& \frac 1{(r+s)} \sum_{n=0}^\infty r^n h(s,-\lambda)^n\cr
&=& \frac 1{(r+s)[1- rh(s,-\lambda)] }\label{eq:Ksl2}
\eea
where $h(s,\lambda)$ is given by Eq.~\eqref{eq:hsl}. 
As mentioned already, it can computed exactly for small values of $r, s$ and is given by Eq.~\eqref{eq:hsl_F}.

Next we calculate the moments of the resetting current using Eqs.~\eqref{eq:Ksl2} along with Eq.~\eqref{eq:hsl_F}. First, we have the Laplace transform of the average resetting current,
\bea
\cal L_{t\to s} [\la \Jres(t)\ra] &=& \frac{\id}{\id \lambda} K(s,\lambda)\bigg|_{\lambda=0}\cr
&=& -\frac{r}{2s^2\sqrt{r+s}}.
\eea
The inverse transform can be performed exactly to obtain, 
\bea
\la \Jres(t)\ra = - \frac 1{2\sqrt{r}}\bigg[(rt-\frac 12) \text{erf}(\sqrt{rt})+ \sqrt{\frac{rt}\pi}e^{-rt}\bigg].~~\label{eq:Jreset_av}
\eea
Note that the above expression is expected to be valid 
for small values of $r \ll 1$ and large $t \gg 1.$
Equation~\eqref{eq:Jreset_av} is very similar to Eq.~\eqref{eq:Jd_av2}, which gives the average diffusive current $\la \Jd(t)\ra$, except, of course, the fact that the average reseting current is negative. In fact, at very long-times $t \gg r^{-1},$ we see a linear growth in magnitude, 
\bea
\la \Jres(t)\ra = -\la \Jd(t)\ra \simeq - \frac {\sqrt{r}t}2,
\eea
At short-times, however, a different behaviour is seen. From Eq.~\eqref{eq:Jreset_av}, for $t \ll r^{-1},$ we have,
\bea
\la \Jres(t)\ra = -\frac{2r t^{3/2}}{3 \sqrt \pi}+ \cal O(t^{5/2}).
\eea
Clearly, at short-times, the resetting current grows much faster than the diffusive current. Figure \ref{fig:Jreset}(a) shows a plot of $\la \Jres(t)\ra$ as a function of $t$ for different values of $r$ which illustrates these features.

It is also interesting to look at the fluctuations of $\Jres.$ From Eq.~\eqref{eq:Ksl2} we can find the Laplace transform of the second moment,
\bea
\cal L_{t\to s} [\la \Jres^2(t)\ra]&=& \frac{\id^2 }{\id \lambda^2} K(s,\lambda)\bigg |_{\lambda=0}\cr
&=& \frac{br}{2 s^2 \sqrt{r+s}}+ \frac{r(\pi r+2s)}{2 \pi s^3(r+s)}
\eea
where, as before, we have used $b= 1- \frac 1{\sqrt 2}.$
Once again, the Laplace transform can be inverted exactly and yields, for $r \ll 1$ and $t \gg 1,$
\bea
\la \Jres^2(t)\ra &=& \frac 1{4 \pi r}\bigg[2e^{-rt}(2-\pi + br \sqrt{\pi t}) + \pi -r + 4 rt \cr
 &+& \pi(rt-1)^2 + b \pi \sqrt{r}(2rt-1)\text{erf}(\sqrt{rt})\bigg].\label{eq:Jreset_sqav}
\eea
The variance of the resetting current $\sigma_\text{reset}^2(t)=\la \Jres^2(t)\ra -\la \Jres(t)\ra^2$ can be computed from Eqs. \eqref{eq:Jreset_sqav} and \eqref{eq:Jreset_av} and it turns out that the variance also increases linearly at the long time limit $t \gg r^{-1}$. In fact, it is straightforward to show that, in this limit, $\sigma_\text{reset}^2(t)=\sigma_\id^2(t)$ [see Eq.~\eqref{eq.sigma_J_d}]. Figure \ref{fig:Jreset}(b) shows $\sigma_\text{reset}^2(t)$ for different values of $r$ obtained from numerical simulations together with the analytical prediction for small $r.$

We conclude the discussion about the resetting current with a brief comment about the probability distribution $\mathscr P(\Jres,t).$ Since $\Jres,$ similar to $\Jd$,  is also a sum of a set of independent variables $J_0(t_i)$, we can use the central limit theorem to predict the behaviour of the corresponding distribution. In fact, for $rt \gg 1,$ one can expect that $\mathscr P(\Jres,t)$ is similar to $P(\Jd,t)$ and has a Gaussian behaviour around the mean value,
\bea
\mathscr P(\Jres,t) \simeq \frac 1{\sqrt{2 \pi \sigma^2_\text{reset}(t)}}\exp\bigg[-\frac{(\Jres-\la \Jres(t)\ra)^2}{2 \sigma_\text{reset}^2(t)}\bigg].\n
\eea

\subsection{Total current}\label{sec:J_tot}

In this section we investigate the behaviour of the total current $J_r,$ as defined in Eq.~\eqref{eq:Jtot_def}. $J_r(t)$ measures the net number of particles which have crossed the central bond towards right (by hopping, or due to resetting) up to time $t.$ As already mentioned, $J_r$ is set to zero after every resetting event; the contribution to the total current comes only from the diffusion of the particles after the last resetting event. 
Consequently, one can write a renewal equation for $P_r(J_r,t),$ the probability that, at time $t,$ the total current will have a value $J_r,$  
\bea
P_r(J_r,t) = e^{-r t} P_0(J_r,t) + r \int_0^t \id s~ e^{-rs} P_0(J_r,s). \label{eq:PJtot_renewal}
\eea
Here $P_0(J_r,s)$ denotes the probability that,  starting from 
$\cal C_0,$ in absence of resetting, $J_r$ number of particles cross the central bond until time $s.$ We will use the above equation to explore $P_r(J_r,t),$ but, first it is useful to investigate the mean and the variance of the total current.

\begin{figure*}
 \centering
  \includegraphics[width=11.4 cm]{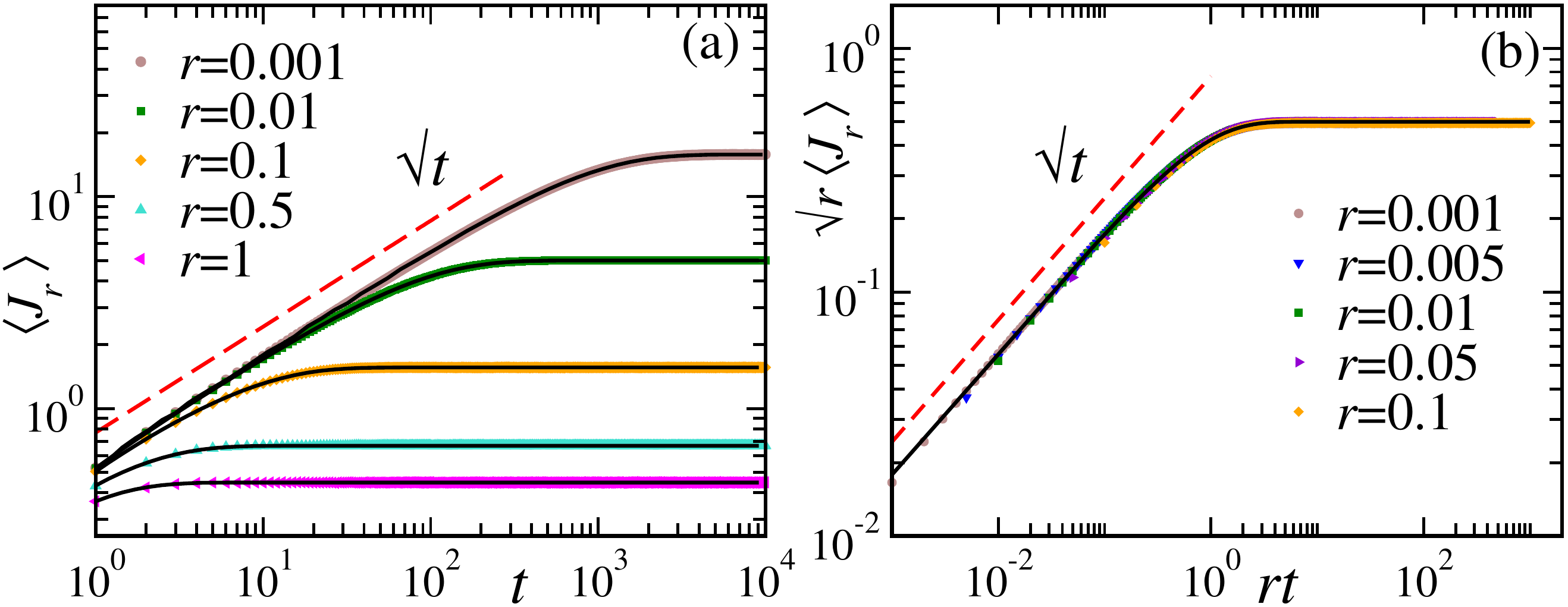}   \includegraphics[width=5.8 cm]{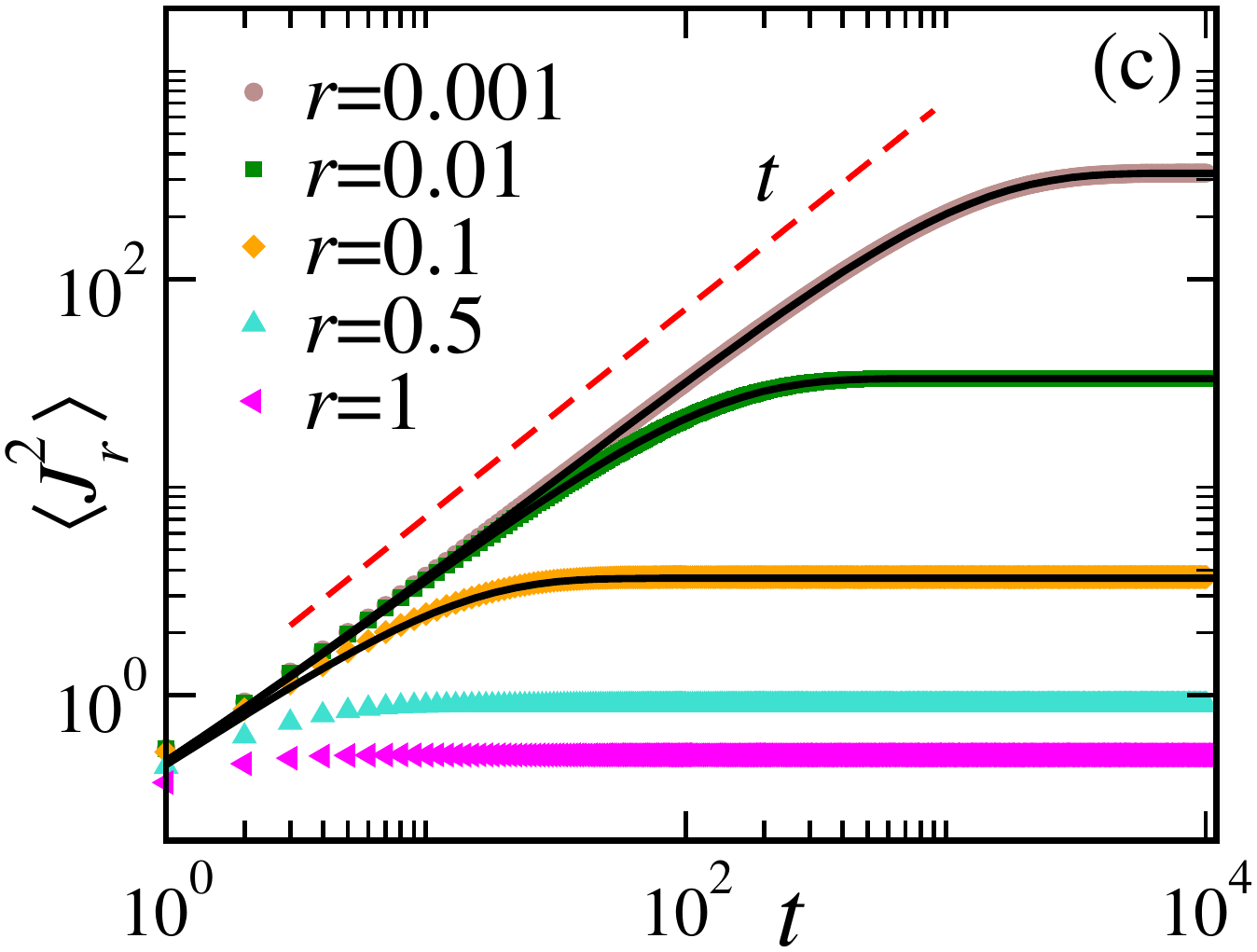} 
 % reset_profile.pdf: 0x0 px, 300dpi, 0.00x0.00 cm, bb=
 \caption{Behaviour of the total current $J_r:$ (a) Plot of average total current $\la J_r(t)\ra$ as a function of time for different values of $r,$ with the uppermost curve corresponding to the smallest value of $r.$ Solid lines and symbols correspond to the analytical result and data from numerical simulations, respectively. (b) Plot of  $\sqrt r \la J_r(t)\ra$ as a function of the scaled variable $rt,$ for small values of $r.$  The solid line corresponding to the scaling function $\text{erf}(\sqrt{rt})/2$ [see Eq.~\eqref{eq:Jtot_scaling}]
   (c) Second moment of the total current $\la J^2_r(t) \ra$ as a function of time $t$ for different values of $r.$ The curves corresponding to small values of $r$ (three upper curves) are compared with the analytical result Eq.~\eqref{eq:Jrsq_t} (solid lines).  A lattice of size $L=1000$ is used for all the numerical simulations.}
 \label{fig:Jtot_r}
\end{figure*}

It is easy to see that all moments of $J_r$ should also satisfy a renewal equation similar to Eq.~\eqref{eq:PJtot_renewal}. In particular, the average total current must satisfy, 
\bea
\la J_r(t) \ra = e^{-rt} \la J_0(t)\ra + r\int_0^t \id \tau ~e^{-r\tau} \la J_0(\tau) \ra \label{eq:Jtot}
\eea
where $\la J_0(t) \ra$ is the average current in absence of resetting, and is  given by Eq.~\eqref{eq:J0_t}. 
%In writing this equation we have used the fact that in absence of resetting diffusive current is the total current. 
Unfortunately, the above integral in Eq.~\eqref{eq:Jtot} cannot be computed analytically. However, it is possible to numerically evaluate the integral and get $\la J_r (t)\ra$ for any time $t.$ 
This is shown in Fig.~\ref{fig:Jtot_r} for different values of $r$ and compared with numerical simulations (symbols) which matches perfectly at all times.

For small values of $r,$ a more explicit expression for the average total current $\la J_r(t) \ra$ can be derived. In that case, it is convenient to rewrite Eq.~\eqref{eq:Jtot} as
\bea
\la J_r(t) \ra= e^{-rt} \la J_0(t) \ra + \int_0^{rt} \id u ~ e^{-u} \left\la J_0 \left(\frac u r \right) \right\ra
\eea
The integral is dominated by the contribution from small $u \sim \cal{O}(1);$ consequently, $u/r$ is large for small $r,$ and we can use the asymptotic expression $\la J_0(u/r) \ra \simeq \sqrt{u/r \pi}.$ Substituting that in the above equation, and performing the integral, we get, for large $t,$
\bea
\la J_r (t) \ra = \frac 1{2\sqrt{r}} \text{erf}(\sqrt{rt}). \label{eq:Jtot_scaling}
\eea
Equation \eqref{eq:Jtot_scaling} provides an explicit expression for the average total current for small $r,$ and in the large time regime. Note that, $\la J_r(t)\ra$ given by the above equation is same as $\la \Jd(t) \ra + \la \Jres(t)\ra,$ as clearly seen from Eqs.~\eqref{eq:Jd_av2} and \eqref{eq:Jreset_av}. This is expected as the total current is a sum of the diffusive current and the resetting current [see Eq.~\eqref{eq:Jtot_def}]. 

We have also measured the total current $J_r$ from numerical simulations. Figure \ref{fig:Jtot_r}(b) shows a plot of $\sqrt{r}\la J_r (t) \ra$ as a function of $rt$ for different (small) values of $r,$ as obtained from numerical simulation; the solid line corresponds to $\text{erf}(\sqrt{rt}).$ The perfect collapse of all the curves verifies our analytical prediction.

From Eq.~\eqref{eq:Jtot_scaling} it can be seen that for $t \ll r^{-1}$ the average total current grows as $\sqrt{t},$ which is a signature of the ordinary SEP. On the other hand, in the large time limit $\la J_r \ra$ reaches a stationary value  $1/2 \sqrt{r}.$ 

In fact, the stationary value of the average total current  $\la J_r \ra$ can be calculated exactly from Eq.~\eqref{eq:Jtot} for any value of $r.$ As we have already seen, at large times $t,$ $\la J_0(t) \ra \sim \sqrt{t},$ hence, the first term in Eq.~\eqref{eq:Jtot} decays exponentially and the large-time behaviour of the average total current is dominated by the second integral in the above equation. Recalling Eq.~\eqref{eq:J0_t} and using the series expansion of the Modified Bessel functions $I_0$ and $I_1,$ (see Ref.~\cite{polylog}, Eq.~10.25.2) we have,
\bea
 && \int_0^t \id \tau ~ e^{-r\tau} \la J_0(\tau)\ra \cr
 &=& \sum_{m=0}^{\infty}\frac1{m!(m+1)!}\frac 1{(r+2)^{2m+3}}\bigg[\Gamma_{2m+3}-\Gamma_{2m+3}\big[(r+2)t\big]\cr
 && +~ (r+2)(m+1)\bigg(\Gamma_{2m+2}-\Gamma_{2m+2}[(r+2)t]\bigg) \bigg],
\eea
where $\Gamma_n$ and $\Gamma_n(x)$ are the Gamma function and the incomplete Gamma function, respectively. $\Gamma_n(x)$ decays to zero for large $x$ for all values of $n,$ and hence, in the  long time limit we have the contributions only from the $t$-independent terms,
\bea
\lim_{t \to \infty}\la J_r \ra &\simeq& r\sum_{m=0}^\infty \frac {(2m+1)!}{m!(m+1)!} \frac {(m+1)(r+2)+2m+2}{(r+2)^{2m+3}}\cr
&=& \frac 1{\sqrt{r(r+4)}}. \label{eq:Jtot_larget}
\eea
Clearly, in the long-time limit the average total current reaches a stationary value $\mu_r = 1/{\sqrt{r(r+4)}}$ which decreases as the resetting rate $r$ is increased. For small $r,$  $\mu_r \approx \frac 1{2\sqrt{r}}$ which is same as what we obtained by taking $t \to \infty$ limit in Eq.~\eqref{eq:Jtot_scaling}. On the other hand, for large $r \gg 1,$ $\mu_r$ approaches $1/r.$

Physically, the limiting behaviours of the stationary value of the average total current can be understood from the following argument. Since the value of $J_r$ is reset to zero after each resetting event, the final contribution to $J_r$ comes only from the diffusion of particles after last resetting event. Moreover, the typical duration since the last resetting event is $\tau_r \sim 1/r.$ For small $r,$ this typical duration is long, and the average diffusive current (without resetting) during this period is  $\sim \sqrt{\tau_r} = 1/\sqrt{r}.$ On the other hand, for large $r,$ $\tau_r$ is small, the diffusive current is $\sim \tau_r = 1/r.$

Next we calculate the second moment of the total current $\la J_r(t)^2 \ra.$ As mentioned already, the second moment also satisfies a renewal equation of the form,
\bea
\la J_r(t)^2 \ra = e^{-rt} \la J^2_0(t)\ra + r \int_0^t \id \tau~ e^{-r \tau} \la J^2_0(\tau)\ra. \label{eq:Jr_sq}
\eea
The above equation is valid at all times and for all values of $r.$ Unfortunately, however, the behaviour of $\la J^2_0(\tau)\ra$ is known only for long time $\tau$ (see Eq.~\eqref{eq:J0_var}) so we are not able to calculate an exact analytical expression for $\la J_r(t)^2 \ra$ for any arbitrary time $t.$   Nevertheless, one can use Eq.~\eqref{eq:Jr_sq} along with Eqs.~\eqref {eq:J0_larget} and \eqref{eq:J0_var} to calculate $\la J_r(t)^2 \ra$ for small values of $r,$ where the integral in  
Eq.~\eqref{eq:Jr_sq} is dominated by the contribution from large $\tau \gg r^{-1}.$ This exercise leads to a simple analytical formula for the second moment for small $r$ (and large time $t$),
\bea
\la J^2_r(t)\ra \simeq \frac 1{\pi r}\left(1 - e^{-rt} \right)+ \frac 1{2\sqrt{r}} \left(1- \frac 1{\sqrt 2}\right) \text{erf}(\sqrt{r t}) \qquad \label{eq:Jrsq_t}
\eea

Figure \ref{fig:Jtot_r}(c) shows the plot of $\la J_r^2(t) \ra$ as a function of $t$ for different values of $r.$ The curves for small $r$ are compared with the analytical result Eq.~\eqref{eq:Jrsq_t} which show an excellent match. Similar to the average total current, the second moment $\la J_r^2(t) \ra$ also eventually reaches a stationary value which, for small values of $r,$ can be obtained by taking $\lim t \to \infty$ in Eq.~\eqref{eq:Jrsq_t}, 
\bea
\la J_r^2 \ra = \frac 1 {\pi r} + \frac 1{2\sqrt{r}} \left(1- \frac 1{\sqrt 2}\right). \label{eq:Jtotsq_larget}
\eea
One can immediately calculate the stationary value of the variance $\sigma_r^2 = \la J_r^2 \ra - \la J_r \ra^2;$ as $r \to 0,$ $\sigma_r^2 \simeq (4- \pi)/ 4 \pi r.$ 

On the other hand, for short time $t,$ we have, 
\bea
\la J_r^2 (t)\ra \simeq \frac t \pi + \sqrt{\frac t \pi }\left(1- \frac 1{\sqrt 2}\right)\label{eq:Jsq_shortt}
\eea
At very short-times, one expects a $\sqrt{t}$ behaviour which crosses over to a linear behavior as $t$ is increased. This is also seen in Figure \ref{fig:Jtot_r}(c),  where the approach to the stationary value appears predominantly linear.\\

\begin{figure}[t]
 \centering
 \includegraphics[width=6.5 cm]{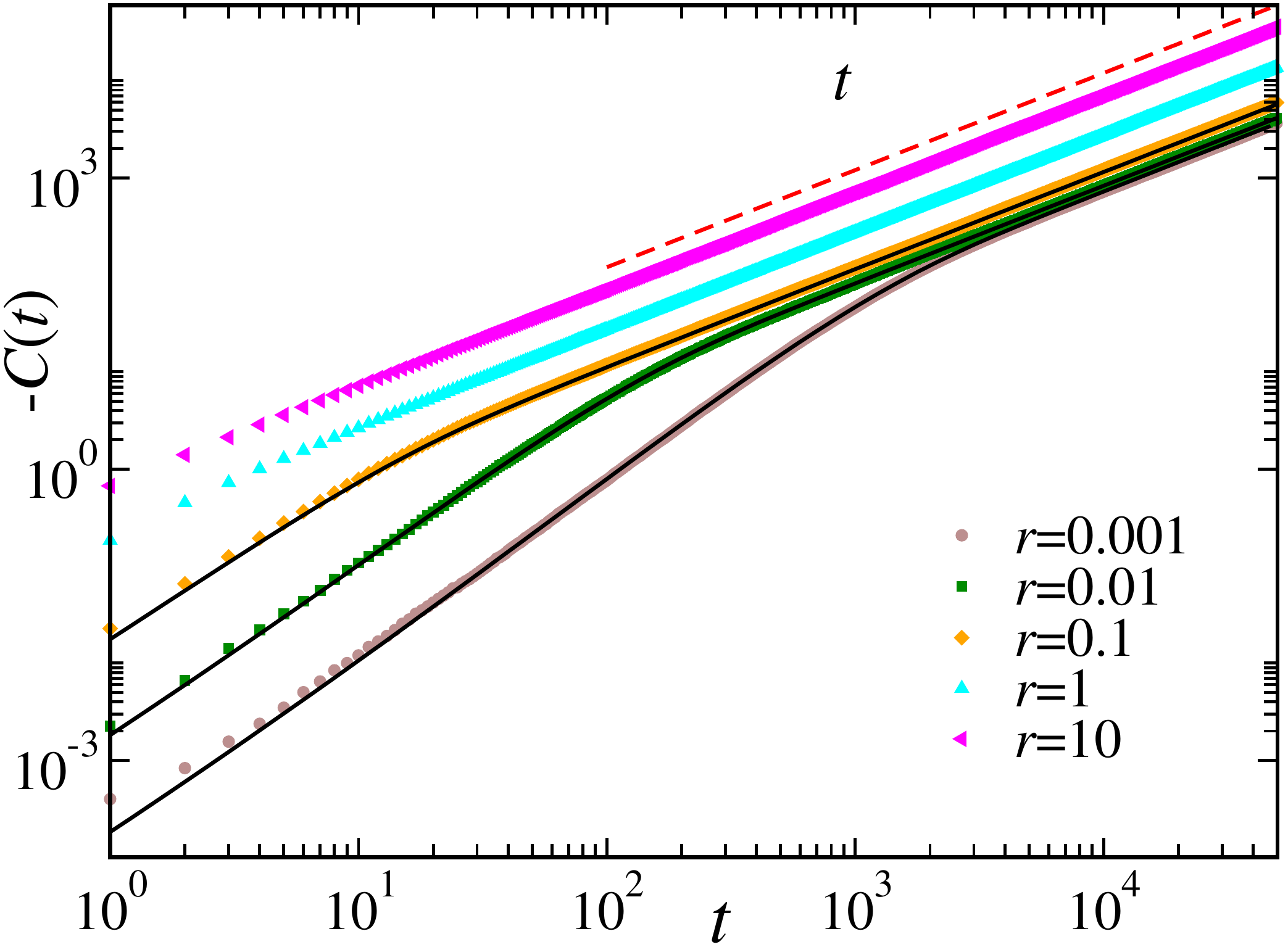}
 % corrJ.pdf: 0x0 px, 300dpi, 0.00x0.00 cm, bb=
 \caption{Correlation between diffusive and resetting currents: Plot of $-C(t)$ as a function of time $t$ for different values of $r;$ the lowest curve corresponds to the smallest value of $r.$ The data from the numrical simulations (symbols) are compared to the analytical prediction from Eq.~\eqref{eq:Ct} (solid lines) for small values of $r.$ Lattice size $L=1000$ is used for the simulations.}
 \label{fig:J_cor}
\end{figure}

\noindent {\bf Correlation between $\Jd$ and $\Jres$:} 
The computation of the second moment of the total current 
$J_r$ provides a way to estimate the correlation between the diffusive and resetting components of the current. 
From the definition of the total current Eq.~\eqref{eq:Jtot_def}, we get,
\bea
\la J_r^2(t) \ra = \la \Jd^2(t) \ra + \la \Jres^2(t) \ra + 2 \la \Jd(t)\Jres(t) \ra. 
\eea
The connected correlation $C(t)= \la \Jd(t) \Jres(t) \ra - \la \Jd(t)\ra \la \Jres(t) \ra$ is then given by,
\bea
 C(t)= \frac 12 \bigg[\sigma_r^2(t) - \sigma_\id^2(t) - \sigma_\text{reset}^2(t)\bigg]
\eea
where $\sigma_r^2,\sigma_\id^2$ and $\sigma_\text{reset}^2$ are the variances of the total, diffusive and resetting currents, respectively. Using Eqs. ~\eqref{eq:Jrsq_t}, \eqref{eq:Jdsq_av}, and \eqref{eq:Jreset_sqav} along with Eqs.~\eqref{eq:Jtot_scaling}, \eqref{eq:Jd_av2} and \eqref{eq:Jreset_av}, we get, for small values of $r,$
\bea
C(t) &=&\frac 1{4 \pi r}\Bigg[4-\pi + rt e^{-2rt}\cr
&+& e^{-rt}\left(\pi -4 -2 r\sqrt{\pi t}[b-\sqrt{r}t~ \text{erf}(\sqrt{rt})]\right)\cr
&-& rt(4-\pi+\pi rt)-b\pi\sqrt{r}(2rt-1)\text{erf}(\sqrt{rt}) \cr
&+& \pi \left(r^2t^2-\frac 14\right)\text{erf}(\sqrt{rt})^2\Bigg]. \label{eq:Ct}
\eea
Clearly, the diffusive and resetting currents are strongly correlated. To understand the nature of this correlation we look at the limiting behaviour of $C(t).$ 
At long times $t \gg r^{-1},$ we get a linear temporal growth from  Eq.~\eqref{eq:Ct},
\bea
C(t) \simeq - \sigma_\id^2(t) \simeq -t \left[\frac{4-\pi}{4\pi} + \frac {b\sqrt{r}}2\right].
\eea
On the other hand, for short-times $t \ll r^{-1}$ (but $t \gg 1$) we get,
\bea
C(t) = - \frac{2 b r}{3 \sqrt{\pi}}t^{3/2} + \cal O(t^2).
\eea
In fact, the correlation remains negative at all times. Figure~\ref{fig:J_cor} shows a plot of $-C(t)$ {\it vs} $t$ for different values of $r$ obtained from numerical simulations (symbols) along with the analytical prediction (solid lines) for small values of $r.$

The presence of a non-trivial correlation between the diffusive and resetting currents suggests that even though the fluctuations of both these components of current are Gaussian in nature, the distribution of the total current need not be so. In the following we investigate this issue and show that, indeed the fluctuations of $J_r$ are characterized by a strongly non-Gaussian distribution.\\

\begin{figure}[t]
 \centering
 \includegraphics[width=8.8 cm]{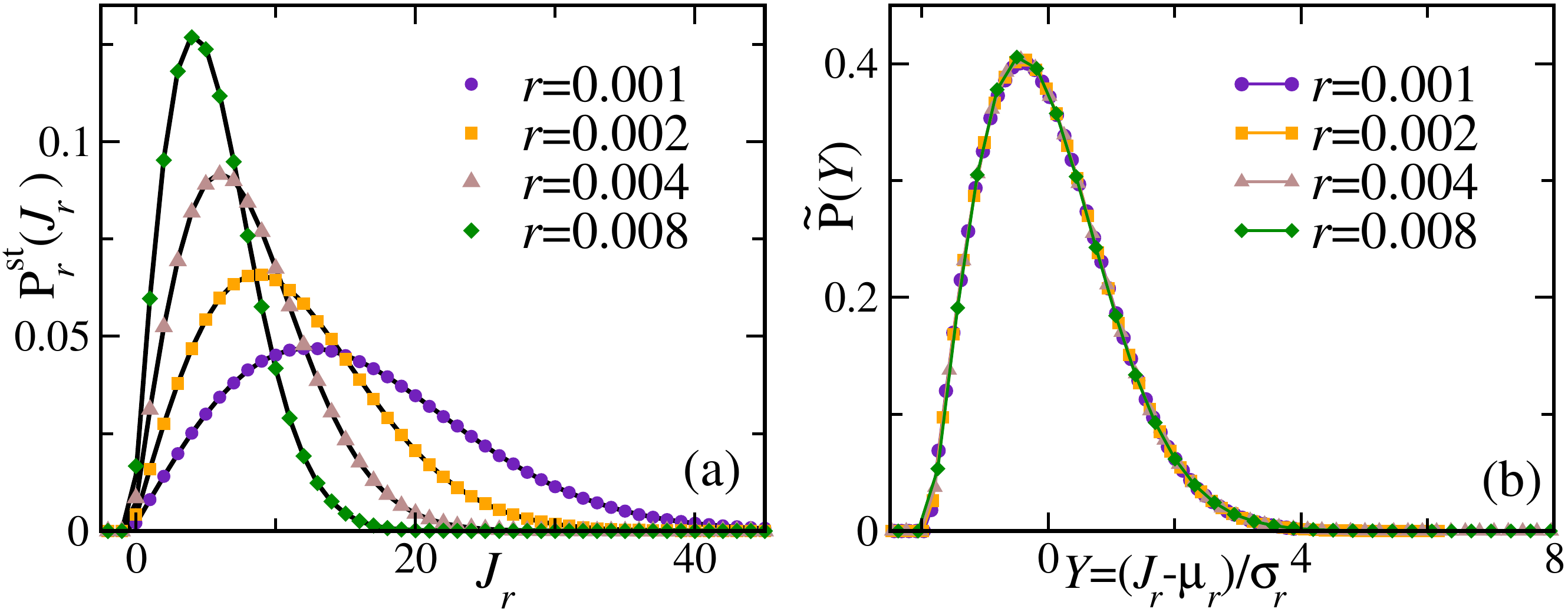}
 % traj.pdf: 0x0 pixel, 300dpi, 0.00x0.00 cm, bb=
 \caption{Stationary probability distribution of the total current $P_r^\text{st}(J_r)$: (a) Plot of $P_r^\text{st}(J_r)$  for different (small) values of $r;$ symbols represent the data obtained from numerical simulations and solid lines correspond to the analytical result obtained using Eq.~\eqref{eq:P_Jrst}. (b) Plot of the same data as in (a) as a function of $Y=(J_r - \mu_r)/\sigma_r$. For simulations we have used $L=1000.$}\label{fig:P_Jtot}
\end{figure}

\noindent {\bf Probability distribution of $J_r$: } In this Section we explore the behaviour of the probability distribution of the total current $P_r(J_r,t)$ using  the renewal equation~\eqref{eq:PJtot_renewal}. 
%$P_r(J_r,t)$ satisfies a renewal equation which can be used to obtain the said distribution given the distribution in absence of resetting. 
In the absence of resetting, the fluctuations of the total (diffusive) current are characterized by a Gaussian distribution in the long-time limit (see Appendix \ref{sec:J0_dist} for more details). Using the Gaussian form of $P_0(J_r,t)$ one can calculate the total current distribution $P_r(J_r,t)$ for small values of $r$ (for small $r$ the integral is dominated by the large $t$ contribution). It is particularly interesting to look at the stationary distribution,

\bea
P_r^\text{st}(J_r) = r \int_0^\infty \id \tau  \frac {e^{- r \tau}}{\sqrt{2 \pi \sigma_\tau^2}}\exp{\bigg[-\frac{(J_r - \mu_\tau)^2}{2 \sigma_\tau^2}\bigg]} \label{eq:P_Jtot}
\eea
where $\mu_\tau = \sqrt{\tau/\pi}$ and $\sigma_\tau^2=  \sqrt{\tau/\pi}(1-1/\sqrt{2})$ are the mean and the variance of the current in absence of resetting, respectively. Clearly, for any finite value of $J_r,$ the Gaussian part of the integrand, \ie, $\exp{[-(J_r-\mu_\tau)^2/2\sigma_\tau^2]},$ vanishes both at $\tau \to 0$ and $\tau \to  \infty$ limits, ensuring that the integral is convergent.  One can then use the series expansion of $e^{-r\tau}$ in Eq.~\eqref{eq:P_Jtot} to express $P_r^\text{st}(J_r)$ as an infinite sum of integrals,
\bea
P_r^\text{st}(J_r) = r \sum_{n=0}^{\infty} \frac{(-r)^n}{\sqrt{2\pi}n!} \int_0^\infty \id \tau \frac{\tau^n}{\sigma_\tau}\exp{\bigg[-\frac{(J_r - \mu_\tau)^2}{2 \sigma_\tau^2}\bigg]}.\n
\eea
Because of the asymptotic properties of $\exp{[-(J_r-\mu_\tau)^2/2\sigma_\tau^2]}$ mentioned above, each of these integrals converge. It turns out that, these integrals can be evaluated exactly for all values of $n$ and yields an explicit expression for the stationary distribution  in the form of an infinite series,
% 
% \added{From the small and large $\tau$ asymptotic behaviour of the integrand, it is easy to realise that
% the above integral is convergent for any given $J_r.$ We
% use the series expansion of $e^{-r \tau}$ in Eq.~\eqref{eq:P_Jtot} to evaluate $P_r^\text{st}(J_r)$  as a sum of integrals. Each integral in the sum converges due to the reason mentioned above, and we get an explicit expression for $P_r^\text{st}(J_r)$ as a series sum,}
\bea
P_r^\text{st}(J_r) &=& \frac{2 \sqrt{2} r}{\pi^{1/4}\sqrt{b}} \exp{\bigg(\frac {J_r}b\bigg)} \cr
&& \times \sum_{n=0}^{\infty} \frac{(-r)^n}{n!} \big(\sqrt \pi J_r\big)^{2n+\frac 32}K_{2n+\frac 32}\bigg(\frac {J_r}b\bigg).\label{eq:P_Jrst} 
\eea
We here have used $b=(1-1/\sqrt{2})$ for brevity, and  $K_\nu(z)$ is the Modified Bessel function of the second kind \cite{polylog} (see Eq.~10.31.1 therein). Convergence of the original integral in Eq.~\eqref{eq:P_Jtot} ensures that the series is also convergent for any finite $J_r.$ Hence, the stationary distribution $P_r^\text{st}(J_r)$ can be computed to arbitrary accuracy using Eq.~\eqref{eq:P_Jrst}. This is demonstrated in Fig. \ref{fig:P_Jtot}(a) where the theoretical computation is plotted together with the simulation results. 
%The excellent agreement between the two validates our prediction.
%The distribution $P_r^\text{st}(J_r)$ computed thus is compared to the same obtained from numerical simulations in Fig. \ref{fig:P_Jtot}(a) which show an excellent match.

% 
% The stationary value of the variance $\sigma_r^2  \equiv  \la J_r^2 \ra - \la J_r \ra^2$ can immediately be calculated  from Eqs.~\eqref{eq:Jtot_larget} and\eqref{eq:Jtotsq_larget} which we will use later.

%Some comments are in order regarding the nature of the distribution of the total current.

The stationary distribution has some interesting features which are visible from Fig.~\ref{fig:P_Jtot}(a). First, it is apparent that $P_r^\text{st}(J_r)$ is vanishingly small for negative values of $J_r.$  This can be understood in the following way. Let us recall that, at any time, the total current is nothing but the net number of particles hopping across the central bond since the last resetting event, \ie,  after being brought to the configuration ${\cal C}_0$ where the left half of the lattice is filled-up. To produce a negative current, the number of particles crossing the central bond from left to right should be lower than that from right to left \ie, there should be a net flux of the particles to the left. Since, the particles are allowed to hop only to empty neighbouring sites, starting from the configuration ${\cal C}_0,$ this is an extremely unlikely event and has a vanishingly small probability.

% To understand this, let us reacll that the total current is nothing but the net number of particles crossing the central bond since the last resetting event. After each resetting, the system is brought back to the configuration $\cal C_0$ where all the sites to the left of the central bond are filled with particles. Hence, it is very unlikely to have a negative current \ie, net flux of particles to the left. 

% This, in fact, is a direct consequence of the choice of the resetting configuration ${\cal C}_0$ where left half of the system is tottaly filled by the particles. The total current is nothing but the net hopping current since the last resetting event,hence, it is very unlikely to have a negative current \ie, net flux of particles to the left. 

Secondly, it also appears that $P_r^\text{st}(J_r)$ is strongly non-Gaussian which is manifest in the asymmetric behaviour of the two tails, as seen in Fig.~\ref{fig:P_Jtot}(a). 
%the decay of the distribution at the two tails, namely near $J_r=0$ and large $J_r,$ are very asymmetric. 
To characterize this asymmetry and the non-Gaussian nature quantitatively we look at the decay of $P_r^\text{st}(J_r)$ at the two tails, namely, near $J_r=0$ and large $J_r.$ Near $J_r=0,$ for small values of $r,$  the behaviour is dominated by the $n=0$ term in  Eq.~\eqref{eq:P_Jrst}. One can then use the asymptotic behaviour of  $K_{\frac 32}(z)$ near $z=0$
% \bea
% K_{\frac 32}(z) \simeq \sqrt{\frac \pi 2} \frac 1{z^{3/2}}
% \eea
to get
\bea
P_r^\text{st}(J_r) \approx 2 \pi r \left (J_r + 1 - \frac 1{\sqrt{2}} \right) + \cal O(r^2). \label{eq:Pjtot_smallj}
\eea
Clearly, for small values of $r,$ the probability distribution of the total current $J_r$ decays linearly near  $J_r=0.$ 
%and vanishes for  $J_r \le J_r^*$ where $J_r^* = \frac 1 {\sqrt{2}} -1 \approx -0.29.$ 

To determine how $P_r^\text{st}(J_r)$ decays for large $J_r$ we use the asymptotic behaviour of $K_\nu(z);$ for large values of the argument $z,$ we have (see Ref.~\cite{polylog}, Eq.~10.40.2),
\bea
\lim_{z \to \infty} K_\nu(z) \approx \sqrt{\frac \pi {2z}} e^{-z}. \label{eq:K_largez}
\eea
Using that in Eq.~\eqref{eq:P_Jrst} and performing the sum over $n,$ we get,
\bea
P_r^\text{st}(J_r) \approx 2 \pi r J_r e^{-\pi r J_r^2} + \cal{O}(r^2). \label{eq:Pjtot_largej}
\eea
Note that the above expression holds true to the leading order in $r,$ higher order corrections can be systematically calculated by including higher order terms in \eqref{eq:K_largez}. 

%it has the following behaviour to the leading order in $r:$ Near $J_r=0,$ the distribution increases linearly,

We conclude the discussion about $P_r^\text{st}(J_r)$  with one final remark. From our numerical data, we observe a surprising collapse of the current distribution when plotted as a function of the scaled variable $Y=(J_r-\mu_r)/\sigma_r$ where $\mu_r$ and $\sigma_r$ are, respectively, the mean and the variance of 
$J_r.$
The collapse is shown in Fig.~\ref{fig:P_Jtot}(b) where the scaled distribution $\tilde P(Y)$  appears to be independent of $r$ as the curves corresponding to different values of $r$ from Fig.~\ref{fig:P_Jtot}(a) fall on top of each other. 
% . 
% To understand the nature of the fluctuations of $J_r$ around its mean $\mu_r$ [given in Eq.~(\ref{eq:Jtot_larget})] we look at the distribution of the scaled variable $Y=(J_r-\mu_r)/\sigma_r$. Figure~\ref{fig:P_Jtot}(b) shows a plot of the corresponding distribution $\tilde P(Y)$  versus $Y;$ remarkably, it appears that 
To understand this collapse, let us look at $\tilde P(Y)$ predicted from Eqs.~\eqref{eq:Pjtot_largej} and \eqref{eq:Pjtot_smallj}. 
Recalling that for small values of $r,$ $\mu_r \simeq \frac1{2\sqrt{r}}$ and $\sigma_r \simeq \sqrt{\frac{(4-\pi)}{4 \pi r}},$  
we get from Eq.~\eqref{eq:Pjtot_largej},
\bea
\tilde P(Y) &\approx& \frac 12 \bigg(\sqrt{\pi(4-\pi)}+ (4-\pi)Y\bigg) e^{-\frac 14 (\sqrt \pi + \sqrt{4-\pi} Y)^2}\cr
&& \qquad + ~ \cal{O}(r^{3/2}).\label{eq:PY_large}
\eea
Clearly, to the leading order in $r,$ $\tilde P(Y)$ calculated from Eq.~\eqref{eq:Pjtot_largej} (corresponding to large values of $J_r$) is independent of $r,$ and is consistent with the scaling collapse observed in Fig.~\ref{fig:P_Jtot}(b). 
On the other hand, it can be easily seen, that Eq.\eqref{eq:Pjtot_smallj} does not lend itself to a similar form; $\tilde P(Y)$ derived from Eq.\eqref{eq:Pjtot_smallj} depends explicitly on $r,$
\bea
\tilde P(Y) &\approx& \frac 12 \sqrt{\pi(4-\pi)}\bigg(1+ \sqrt{\frac{4-\pi}{\pi}}Y + (2- \sqrt{2})\sqrt{r}\bigg) \cr
&& + \cal{O}(r^{3/2}).\label{eq:PY_small}
\eea
Hence, while for large positive $J_r$ ($\gtrsim \mu_r+\sigma_r$), the distribution $\tilde P(Y)$ becomes independent of $r$, it is not the case in the $J_r \to 0$ limit. 
 Indeed, as seen from Eq.~\eqref{eq:PY_small}, $\tilde P(Y)$ explicitly depends on $r.$  However, notice that the 
 $r$-dependence in Eq.~\eqref{eq:PY_small} comes in the form of an additional term proportional to $\sqrt{r},$ which is vanishingly small for $r \ll 1.$  
  %`difference' between the value of the distributions at small $J_r$ for different $r$ is $\sim \sqrt{r}.$ 
This makes the expected mismatch in the collapse at the left tail in Fig.~\ref{fig:P_Jtot}(b) practically invisible 
%  at the scale of Fig.~\ref{fig:P_Jtot}(b) where 
 where an apparent collapse is also observed.
% even at the left tail. 

\section{Conclusion} \label{sec:conc}
%Studying the effect of resetting in stochastic processes is a subject of current interest. While most of the studies concentrated on single particle systems, it is important to ask how introduction of resetting mechanism affects the behaviour of interacting particle systems.

In this article, we explore the effect of stochastic resetting on interacting many particle systems. To this end, we study the dynamical properties of a canonical set-up, namely, the symmetric exclusion process in the presence of stochastic resetting. The resetting is implemented by interrupting the dynamical evolution of the exclusion process with some rate $r$ and restarting it from a step-like configuration where all the particles are clustered together in the left-half of the system. 

We find that the presence of resetting strongly affects the behaviour of the system. The key findings are as follows. First, in a finite size system, the density profile evolves to an inhomogeneous stationary profile in contrast to the flat profile in the  absence of resetting. We have exactly calculated the full time-dependent density profile for arbitrary resetting rate $r.$ Secondly, we find that, in a thermodynamically large system the resetting mechanism drastically changes the $\sqrt{t}$ growth of the diffusive current to linear in $t$.  We have explicitly  computed the mean and variance of the diffusive current, the latter is also shown to have a linear growth in the long-time regime. Apart from the diffusive current, we also identify the another component of the current which  arises due the resetting move and show that this resetting current is negative, with a linear temporal growth in magnitude.   
The moments of the total current, \ie, the sum of the diffusive and resetting current, are also calculated using the renewal approach.

We also have investigated the probability distribution of the diffusive current $\Jd,$ resetting current $\Jres,$ as well the total current $J_r.$ We have found that that while the typical fluctuations of $\Jd$ and $\Jres$ are Gaussian in nature, the distribution of $J_r$ is strictly non-Gaussian. The non-Gaussian nature is manifest in the asymmetric asymptotic behaviour of the distribution at the two tails, which we also demonstrate.

Our study opens up a new direction in the area of stochastic resetting and gives rise to a wide range of further questions. For example, it would be interesting to study the effect of stochastic resetting in other interacting particle systems, e.g., the asymmetric exclusion process, driven and equilibrium lattice gas models etc. Furthermore, it would also be interesting to study behavior of these interacting particle systems under various other resetting mechanisms like resetting at power-law times or time-dependent resetting etc.  

Apart from these theoretical questions, 
the framework of stochastic resetting in exclusion processes can also be relevant in the context of certain biophysical systems. For example, stochastic motion of backtracked RNA polymerases can be modelled as an interacting many particle random walk on the DNA template, with RNA cleavage playing the role of resetting dynamics \cite{bio3,Lisica}.  \\
Similarly, motion of two-headed molecular motors such as kinesin and Myosin V moving on a polymeric track can be modelled as an energy driven hopping process in the presence of backward jumps (or resetting)\cite{Astumian}. We believe that the formalism introduced in the present work will be useful in understanding such systems.

\acknowledgements
The authors acknowledge useful discussions with Christian Maes and Sanjib Sabhapandit. U.B. acknowledges support from Science and Engineering Research Board (SERB), India under Ramanujan Fellowship (Grant No. SB/S2/RJN-077/2018). A. K. acknowledges support from DST 
grant under project No. ECR/2017/000634. A.P. gratefully acknowledges support from  the Raymond and Beverly Sackler Post-Doctoral Scholarship at Tel-Aviv University.

\appendix
%\section{Master equation for SEP under resetting}

\section{Density profile of SEP in absence of resetting}\label{app:sep}
In this section we present a brief account of the dynamical evolution of the density profile and current for ordinary SEP, starting from the step-like configuration $\cal C_0.$ In absence of resetting, the time evolution of the system is governed by the free Markov matrix ${\cal L}_0$ which yields, for the density profile,
\bea
\frac{\id}{\id t} \rho_0(x,t) &=& \rho_0(x+1,t)+ \rho_0(x-1,t)- 2\rho_0(x,t). \;\;
\eea
The corresponding Fourier components $\tilde \rho_0(n,t)$ evolve following,
%Substituing Eq.~\eqref{eq:rho_dft} in Eq.~\eqref{eq:rho0t}, we get,
\bea
\frac{\id }{\id t} \tilde \rho_0(n,t) = - \lambda_n \tilde \rho_0(n,t) 
\eea
where, as before, $\lambda_n= 2\left(1- \cos \frac{2 \pi n}{L}\right),$ with $n=0,1,2,\dots L-1.$ The above equation is  immediately solved to obtain,
\bea
\tilde \rho_0(n,t) = e^{-\lambda_n t} \tilde \phi(n). \label{eq:rho0_n}
\eea
where $\tilde \phi(n)$ corresponds to the initial profile $\phi(x).$ Note that $\lambda_0=0$ and hence $\tilde \rho_0(0,t)= \tilde \phi(0) = \frac L2$ does not evolve with time.

The spatial density profile is obtained by taking inverse Fourier transform of Eq.~\eqref{eq:rho0_n}. In particular, for the step-like initial profile $\phi(x)=1- \Theta(x+1-\frac L2)$ we have,
\bea
\rho_0(x,t) = \frac 12 + \frac 1L \sum_{n=1,3}^{L-1} e^{-i \frac{2 \pi n x}{L}} \left(1+i \cot \frac {\pi n}{L}\right) e^{-\lambda_n t}~~\;\; \label{eq:rho0xt}
\eea

\section{Behaviour of current in absence of resetting}\label{sec:J0}

In the absence of resetting the only source of current in SEP is the hopping dynamics of the particles. The average instantaneous current across the initial step, \ie, across the central bond $\left(\frac L2 -1, \frac L2 \right)$ is given by,
\bea
\la j_0(t) \ra &=& \rho_0\left(\frac L2-1,t\right) - \rho_0\left(\frac L2,t\right) \cr
&=& \frac 2L \sum_{n=1,3}^{L-1} e^{-\lambda_n t} \label{eq:j0t}
\eea
where we have used Eq.~\eqref{eq:rho0xt} to calculate the average densities at the sites $x=\frac L2-1$ and $x=\frac L2.$ Clearly, in the long-time limit $t \to \infty,$ the instantaneous current vanishes as the density profile becomes flat.

We are interested in the time-integrated current $J_0(t) = \int_0^t \id s~ j_0(s)$ which measures the net number of particles crossing the central bond towards right. The average time-integrated current is obtained by integrating Eq.~\eqref{eq:j0t},
\bea
\la J_0(t) \ra =  \frac 2L \sum_{n=1,3}^{L-1} \frac {1}{\lambda_n}(1-e^{-\lambda_n t}).
\eea
For any finite $L,$ the average time-integrated current $J_0(t)$ saturates to an $L$-dependent constant value in the long-time limit. 

To understand the behaviour of a thermodynamically large system, one has to take the limit $L \to \infty$ first. In this case, the sum in Eq.~\eqref{eq:j0t} can be can be converted to an integral by denoting $q= 2 \pi n/L,$ and we have the mean instantaneous current,
\bea
\la j_0(t) \ra &=& \int_0^{2 \pi} \frac{\id q}{2 \pi}~ e^{-2(1-\cos q)t} 
\\ [0.25em] \n
&=& e^{-2 t} I_0(2 t).
\eea
Here $I_0(x)$ is the Modified Bessel function of the first kind \cite{polylog} (see Eq.~10.25.2 therein). In this limit, the average time-integrated current becomes,
\bea
\la J_0(t) \ra = e^{-2t} t [I_0(2t)+I_1(2t)].\label{eq:J0_t}
\eea
For large values of the argument $x,$ both $I_0(x)$ and $I_1(x)$ have the same asymptotic behaviour (see \cite{polylog}, Eq.~10.40.1 	),
\bea
\lim_{x\to \infty}I_{0,1}(2x) \sim  \frac{e^{2x}}{2 \sqrt{\pi x}} \label{eq:I01_large} %\qquad \text{for} \;\; x \gg 1
\eea
which yields, in the long-time regime,
\bea
\la J_0(t) \ra \simeq \sqrt{\frac t \pi}. \label{eq:J0_larget}
\eea
%At long time, for an infinitely large system, the current through the central bond increases as $\sqrt{t}.$ 
This result has been obtained in Ref.~\cite{Derrida1}, albeit using a different method.  In fact, it has also been shown \cite{Derrida1} that, in the long-time regime, all the higher moments of $J_0$ show a similar behaviour. In particular, the variance is given by,
\bea
\la J_0^2(t) \ra - \la J_0(t) \ra^2 \simeq \sqrt{\frac{t}{\pi}}\left(1- \frac 1{\sqrt{2}}\right). \label{eq:J0_var}
\eea
%We use this result to investigate the distribution of the current $J_0$ in the following. \\
The above equation is used in Eq.~\eqref{eq:Jr_sq} in the main text to calculate $\la J_r^2\ra.$

\subsection {Probability Distribution of $J_0$} \label{sec:J0_dist}

For ordinary SEP, the probability distribution of the time-integrated current $J_0$  was explored in Ref.~\cite{Derrida1}. There the authors considered a scenario where, initially, each site to the left (respectively, right) of the origin ($x \le 0$ and $x>0$ respectively) is occupied with probability $\rho_a$ (respectively $\rho_b$). It was shown that, for large $t,$ the moment generating function of the total particle flux $J_0(t)$ through the origin is given by,
\bea
\la e^{\lambda J_0(t)} \ra \sim e^{\sqrt{t} F(\omega)}
\eea
where $\omega = \rho_a(e^\lambda -1) + \rho_b(e^{-\lambda}-1) + \rho_a \rho_b(e^\lambda -1)$ and 
\bea
F(\omega) &=& \frac 1{\sqrt{\pi}} \sum_{n=1}^{\infty} \frac{(-1)^{n+1}\omega^n}{n^{3/2}} \cr
&\equiv& -\frac 1{\sqrt{\pi}} \text{PolyLog}_{3/2}(-\omega).
\eea

In our case, we have $\rho_a=1$ and $\rho_b=0$ which simplifies $\omega$ and in turn $F(\omega);$ we get $\omega = e^{\lambda}-1$ and
\bea
F(\lambda) = -\frac 1{\sqrt{\pi}} \text{PolyLog}_{3/2}(1-e^{\lambda})
\eea
which is quoted in Eq.~\eqref{eq:F_lam} in the main text.

It has been shown in Ref.~\cite{Derrida1}  that the corresponding probability distribution $P_0(J_0,t),$ in the long-time limit, is of the form,
\bea
P_0(J_0,t) \sim e^{\sqrt{t}G(J_0/\sqrt{t})}. \label{eq:PJ0_LD}
\eea
The large deviation function $G(q=J_0/\sqrt{t})$ is related to $F(\lambda)$ through a Legendre transform,
\bea
G(q) &=& \min_{\lambda}[F(\lambda) - \lambda q] \cr
&=& F(\lambda^*) - \lambda^* q, \label{eq:Gq}
\eea
where $\lambda^*$ corresponds to the minimum of the function $F(\lambda)- \lambda q$ and is obtained by solving $\frac{\id F(\lambda)}{\id \lambda} = q.$ It is easy to see that for small values of $q,$ $\lambda^*$ is also small. Hence, it is convenient to
use the series expansion of  $F(\lambda)$ near $\lambda=0$  
\bea
F(\lambda)= \frac \lambda {\sqrt{\pi}} + \frac{\lambda^2}{2\sqrt{\pi}}\left(1- \frac 1{\sqrt{2}}\right) + \cal O(\lambda^3),
\eea
to find $\lambda^*$ for small values of $q.$ Restricting ourselves to the quadratic order in $\lambda,$ we get 
$\lambda^*=\frac{(q \sqrt{\pi}- 1)\sqrt{2}}{(\sqrt{2}-1)}.$ Substitution of this $\lambda^*$ in Eq.~\eqref{eq:Gq} yields,
\bea
G(q) = \frac{(q - \frac 1 {\sqrt{\pi}})^2}{\frac 2 {\sqrt \pi}(1- \frac 1{\sqrt{2}})}
\eea
Using the above $G(q)$ in Eq.~\eqref{eq:PJ0_LD} results in a Gaussian form for the current distribution,
\bea
P_0(J_0,t) = \frac 1{\sqrt{2 \pi \sigma_0^2(t)}} \exp{\bigg(- \frac{[J_0 - \mu_0(t)]^2}{2 \sigma^2_0(t)}\bigg)}\label{eq:P_J0}
\eea
where the prefactor is just a normalization constant. Here, $\mu_0(t)=\sqrt{\frac t{\pi}}$ is nothing but the average hopping current $\la J_0(t)\ra$ and $\sigma_0^2(t)=\sqrt{\frac{t}{\pi}}\left(1- \frac 1{\sqrt{2}}\right)$ is the variance [see Eq.~\eqref{eq:J0_var}]. Note that, this Gaussian distribution is expected only in the long-time limit, as Eq.~\eqref{eq:PJ0_LD} holds true in this limit only.

\begin{figure}[t]
\vspace*{0.5 cm}
 \centering
  \includegraphics[width=8.8 cm]{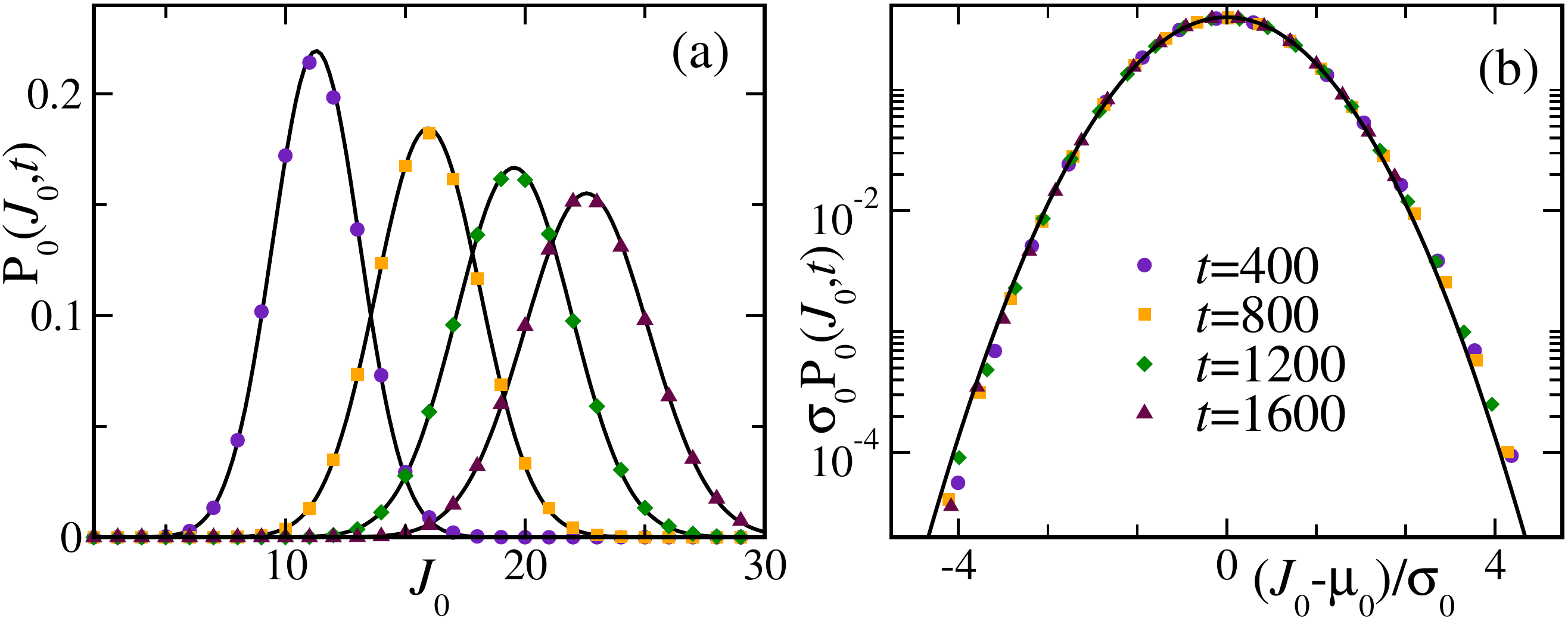}
 % reset_profile.pdf: 0x0 px, 300dpi, 0.00x0.00 cm, bb=
 \caption{(a) Plot of $P_0(J_0,t)$ {\it vs} $J_0$ for different (large) values of $t.$ The symbols indicate the data obtained from numerical simulation of a system of size $L=1000,$ whereas the solid black lines correspond to the Gaussian distribution (see Eq.~\eqref{eq:P_J0}). (b) The same data plotted as function of $(J_0-\mu_0(t))/\sigma_0(t);$ the solid line indicates the standard normal distribution.}
 \label{fig:J0_dist}
\end{figure}

Figure \ref{fig:J0_dist} (a) shows a comparison of $P_0(J_0,t)$ obtained from numerical simulations (symbols) with the that predicted from Eq.~\eqref{eq:P_J0} (solid lines) for different (large) values of $t.$ 
Fig.~\ref{fig:J0_dist} (b) shows the same data but plotted against the scaled variable  $y=\frac{J_0 - \mu_0(t)}{\sigma_0(t)};$ the solid line corresponds to the standard normal distribution $\frac 1{\sqrt{2 \pi}}e^{-y^2/2}.$ The numerical data shows a very good match with the predicted Gaussian curve for typical values of $J_0,$ there are deviations only at the regime $|y| \gg 1$ which are visible only at a logarithmic scale. The large deviation function calculated in Ref~\cite{Derrida1} describes the distribution for these atypical values. However, as shown in Sec. \ref{sec:J_tot}, for our purposes it suffices to consider the typical fluctuations and we use the Gaussian distribution \eqref{eq:P_J0} to calculate the distribution of the diffusive current $\Jd$ in presence of resetting.

\end{document}